\documentclass{article}

\usepackage{graphicx}

\newlength{\bredde}
\def\slash#1{\settowidth{\bredde}{$#1$}\ifmmode\,\raisebox{.15ex}{/}
\hspace*{-\bredde} #1\else$\,\raisebox{.15ex}{/}\hspace*{-\bredde} #1$\fi}
\begin{document}
\topmargin -1.4cm
\oddsidemargin -0.8cm
\evensidemargin -0.8cm
\title{Random Matrix Theory and Quantum Chromodynamics}

\author{Gernot Akemann~\\~\\
Department of Physics,  Bielefeld University, Postfach 100131, D-33501 Bielefeld, Germany}

\maketitle

\begin{abstract}
\noindent
These notes are based on the lectures delivered at the Les Houches Summer School in July 2015. They are addressed at a mixed audience of physicists and mathematicians with some basic working knowledge of random matrix theory. The first part is devoted to the solution of the chiral Gaussian Unitary Ensemble in the presence of characteristic polynomials, using orthogonal polynomial techniques. 
This includes all eigenvalue density correlation functions, smallest eigenvalue distributions and their microscopic limit at the origin. 
These quantities are relevant for the description of the Dirac operator spectrum in Quantum Chromodynamics with three colours in four Euclidean space-time dimensions. In the second part these two theories are related based on symmetries, and the random matrix approximation is explained. In the last part recent developments are covered including the effect of finite chemical potential and finite space-time lattice spacing, and their corresponding orthogonal polynomials. 
We also give some open random matrix problems.

\date{}


\end{abstract}

\thispagestyle{empty}

\tableofcontents



\section{Introduction and motivation}
\label{intro}

In this short introduction we would like to introduce the two players in the title, Random Matrix Theory (RMT) and Quantum Chromodynamics (QCD), on a superficial level. This gives a motivation why and where it will be beneficial to relate these two seemingly unrelated theories, and what the reader may expect to learn from these notes. 

Let us begin with QCD, the theory of the strong interactions\footnote{I apologise for being very elementary here, my audience was not assumed to know particle or general physics.}. It is part of the standard model of elementary particles that also describes the weak and electromagnetic interactions among all elementary particles. The latter two interactions will however not play any role here. The particle content of QCD are the $N_f$ quarks with masses $m_q$, $q=1,\ldots,N_f$, and the gluons. In nature $N_f=6$ flavours have been observed named up, down, strange, charm, top and bottom quark. In the following we will keep $N_f$ as a free parameter, and often consider only the lightest up and down quark ($N_f=2$) as they constitute the most common particles as proton (p), neutron (n) and pions ($\pi^{\pm,0}$). The quarks interact through the gluons, the carriers of force, with a coupling constant $g_s$. Both come in 3 colours and the interaction is described through a field strength and covariant derivative carrying an $SU(3)$ Lie group structure. QCD is a strongly interacting, relativistic quantum field theory (QFT) which is very difficult. Fortunately we will only need to know a few features and some of its global symmetries that will be described in Section \ref{symm}. For a standard textbook on perturbative QCD we refer for example to \cite{Muta}. 

\begin{figure}[h]
\unitlength1cm
\begin{center}
\begin{picture}(12.2,5.5)
\put(1.0,0.3){\vector(1,0){8}}
\put(9.2,0.2){$\mu$}
\put(1.0,0.3){\vector(0,1){5}}
\put(0.1,5.1){$T$}
\put(0.1,4.3){$T_c$}
\put(7.1,4.0){$\Sigma=0$}
\put(1.0,-0.1){$0$}
\put(2.9,1.8){$\Sigma\neq0$}
\put(3.9,4.25){$\bullet$}
\put(8.0,-0.1){$\mu_c$}
\qbezier[60](1.0,4.4)(4.7,4.3)(4.0,4.3)
\qbezier(4.0,4.3)(7.0,4.2)(8.0,0.3)
\end{picture}
\end{center}
\caption{Schematic phase diagram of QCD as a function of temperature $T$ and quark chemical potential $\mu$, for two massless quark flavours. The chemical potential is proportional to the Baryon density. For massless flavours $m_q=0$ the dashed line corresponds to a second order phase transition, merging with the full line representing a first order transition in a tricritical point.
\label{fig:phasediag}}
\end{figure}
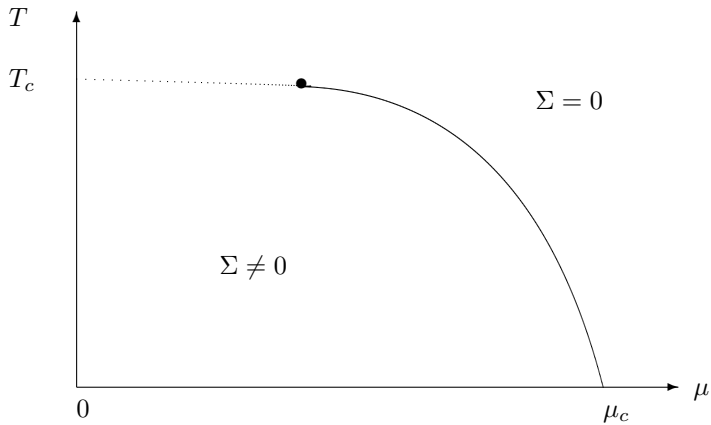

Roughly speaking QCD has two different phases that are schematically depicted in Figure \ref{fig:phasediag}. They are characterised by an order parameter, the chiral condensate $\Sigma$. At high energies corresponding to high temperatures T, $\Sigma=0$ and quarks and gluons form a plasma, that has been observed in collision experiments e.g. at the Relativistic Heavy Ion Collider RHIC in Brookhaven. This phase also existed in the early universe, with the cooling down happening close to  the temperature axis at low Baryon density parametrised by $\mu$. In these situations as well as in single particle collisions produced at collider experiments a perturbative expansion in powers of the coupling $g_s$ typically applies, as described e.g. in \cite{Muta}. The second phase with $\Sigma\neq0$ at low temperature and density is the one in which we live. Here quarks condense into colourless objects, that is into Baryons made of 3 quarks of different colour adding up to white (like p or n), or Mesons made of a quark and an anti-quark with its anti-colour (like the $\pi$'s). Gluons are also confined to these objects, and a proof of confinement of quarks and gluons from QCD is still considered to be an open millennium problem. In QCD it happens that confinement goes along with the spontaneous breaking of chiral symmetry - hence the name of $\Sigma$. This  global symmetry will be described in detail later, and eventually lead us to a RMT description. 
 
In the phase with $\Sigma\neq0$ perturbation theory breaks down and one has to apply other methods. If one wants to continue to work with first principles and the QCD action, one possibility is to study QCD on a finite space-time lattice of volume $V$ and lattice spacing $a$, equipped with a Euclidean metric. The numerical solution of this theory nowadays reproduces the masses of particles that are composed of elementary ones to a very high precision, as testified in the particle data booklet \cite{weblinkPDG}. To that aim two limits have to be made, the continuum limit sending $a\to0$ and the thermodynamical limit sending $V\to\infty$. For a standard textbook on lattice QCD we refer to \cite{MM}. 
A second possibility is to approximate QCD in the confined phase by effective theories, that describe for example only the low energy excitation. One of these is chiral perturbation theory (chPT), see \cite{Scherer} for a review, that describes the low momentum modes that appear after the spontaneous breaking of chiral symmetry, the so-called Goldstone Bosons. Another such theory is RMT. It should be clear by now that RMT will not solve QCD (nor chPT), in the continuum or on the lattice. However, it will describe certain aspects in an analytic fashion, namely the spectral properties of the small eigenvalues of the  QCD Dirac operator $D\!\!\!\!/$. In a finite volume the density of eigenvalues of $D\!\!\!\!/$ satisfies the Banks-Casher relation \cite{BC}
\begin{equation}
\rho_{D\!\!\!\!/}(\lambda\approx0) = \frac{1}{\pi} \Sigma V \ ,
\label{BCrel}
\end{equation}
which relates it to the described setup. In particular RMT will predict the detailed dependence on the parameters $m_q$, $\mu$, $a$, and $V$ after being appropriately rescaled, as well as on a the zero-eigenvalues of $D\!\!\!\!/$ which relate to a topological index $\nu$. And most remarkably these predictions have been verified in comparison to lattice QCD by many groups. We will not repeat these findings here and refer to the literature at the end of each section.

The idea to apply RMT to QCD goes back to the seminal works \cite{SV93,JacIsmail} that started from the simplest case with $N_f=0$ which is called quenched approximation, at $\mu=0=a$. The field has since developed enormously, leading to the detailed analytical knowledge that we will describe. There exist a number of excellent reviews already, \cite{TiloJac,Poul2011}, notably the lecture notes from the Les Houches session in 2004 by J. Verbaarschot \cite{Jac2005}. This brings me to the main goals of these lecture notes, to be addressed in the order of the subsequent sections. What can we predict from RMT that can be compared to lattice QCD? In Section \ref{OPDirac} we will start with the solution of the corresponding RMT using the theory of orthogonal polynomials. This section is rather mathematical and contains no further physics input. The second question, what the limit is, in which QCD reduces to RMT, is addressed in Section \ref{symm}. Here the global symmetries of QCD are explained, leading to chPT and eventually to a RMT description. The more physics inclined reader may jump to this section first. One of my personal motivations why to add another review on this topic is answered in Section \ref{recent} where the development in the last 10 years is reviewed, including some mathematical aspects. This covers the dependence on finite lattice spacing $a$ and on chemical potential $\mu$, for an earlier review on the latter see  \cite{A07}.

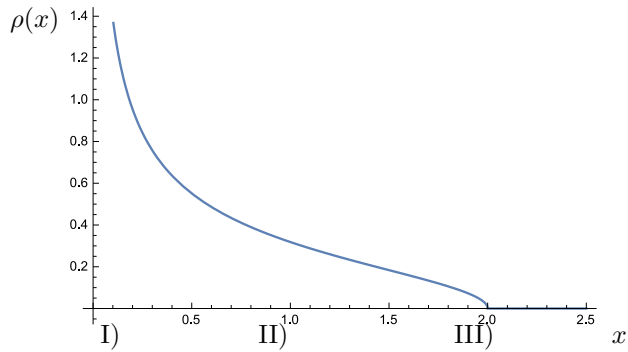
\begin{figure}[b]
\unitlength1cm
\begin{center}
\begin{picture}(12.2,4.5)
\includegraphics[width=7cm]{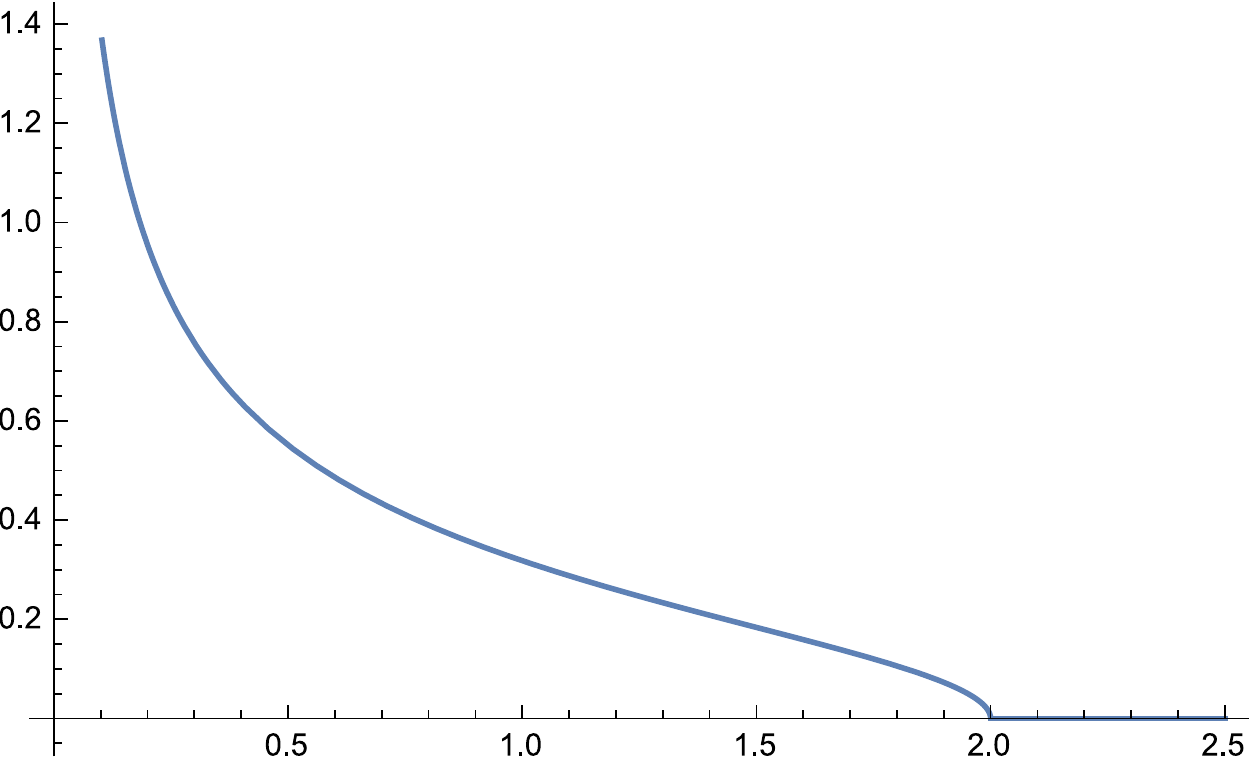}
\put(0.2,-0.2){$x$}
\put(-7.8,4.0){$\rho(x)$}
\put(-1.9,-0.2){III)}
\put(-6.6,-0.2){I)}
\put(-4.5,-0.2){II)}
\end{picture}
\end{center}
\caption{The Marchenko-Pastur law $\rho(x)=\frac{1}{\pi}\sqrt{(2-x)/x}$ which is the limiting global spectral density of the chGUE. We also indicate the locations where different local statistics applies: they are given by I) the Bessel-kernel close to the hard edge, II) the sine-kernel in the bulk and III) the Airy-kernel at the soft edge.
\label{fig:MPdensity}}
\end{figure}

Let us move to RMT, where I assume that the reader has already some working knowledge. RMT is a much older topic, beginning in the late 1920's with Wishart in mathematical statistics and in the late 1950's with Wigner and Dyson in nuclear physics. The number of its applications is huge and still increasing, and we refer to \cite{GMW} for applications to quantum physics and to \cite{Handbook} for a recent compilation containing physics (including QCD in chapter 32), mathematics and more. In these lecture notes we will focus only on the theory of orthogonal polynomials. For the standard Gaussian ensembles they are covered in the last edition of Mehta's classical book \cite{Mehta}, for more recent monographs see \cite{AGZ,BookPeter}. The symmetry class we will be interested in for the application to QCD is the chiral Gaussian Unitary Ensemble (chGUE), which is also known as Wishart or Laguerre Unitary ensemble, and extensions thereof. In the simplest case of $N_f=0$ and when all the abovementioned parameters from QCD are absent it is defined by the probability space of complex $N\times N$ matrices $W$. All its matrix elements are independent and share the same complex normal distribution. The mean density $\rho(x)$ of the positive eigenvalues of $WW^\dag$, where $\dag$ denotes the Hermitian conjugate, is well known in the limit $N\to\infty$. After an appropriate rescaling it is given by the Marchenko-Pastur law depicted in Figure \ref{fig:MPdensity}, cf. the lectures by Bouchaud where this was derived.
The same law is obtained when taking real or quaternion valued matrix elements, constituting the chiral Gaussian Orthogonal and Symplectic Ensemble (chGOE and chGSE), respectively. However, the local statistics of the 3 ensembles differs, and depends on the location in the spectrum we consider. For finite $N$ the eigenvalue statistics follows a determinantal (chGUE) or Pfaffian point process (chGOE, chGSE), and the limiting kernels in the various locations of the spectrum shown in Figure \ref{fig:MPdensity} are given in terms of Bessel, Sine or Airy functions. In view of the relation (\ref{BCrel}) we will be interested in the local statistics for the chGUE at the origin also called hard edge. The other two symmetry classes at the hard edge will also find applications in QFT as pointed out in \cite{Jac3fold}, cf. \cite{TiloJac}, but will not be discussed here. In the limit where RMT applies the eigenvalues $y_j$ of the QCD Dirac operator $D\!\!\!\!/$ will be given by $y_j=\pm\sqrt{x_j}$ and thus come in pairs, where $x_j$ are the eigenvalues of $WW^\dag$ (the positive part $+\sqrt{x_j}$ are the singular values of $W$). After this change of variables the Marchenko-Pastur law becomes a semi-circle, which is why the global density of $D\!\!\!\!/$ at the origin will become constant (and not divergent as in Figure \ref{fig:MPdensity}).
For finite $N$ the chGUE can be solved in terms of the classical Laguerre polynomials.  It will be the content of Sections \ref{OPDirac} and \ref{recent} to include more structure from QCD into this RMT, while maintaining its exact solvability.

\section{Orthogonal polynomial approach to the Dirac operator spectrum}
\label{OPDirac}

In this section we go directly to the RMT of QCD in an eigenvalue representation that generalises the chGUE, by including the mass terms of $N_f$ quark flavours. In Subsection \ref{Defs} we briefly recall the orthogonal polynomial (OP) formalism and define the quantities of our interest. We then attack the mathematical problem of computing these by deriving properties of OP with a general weight, in particular including averages of characteristic polynomials in Subsection \ref{genweight}. This allows us to determine all quantities explicitly in terms of standard Laguerre polynomials in Subsection \ref{allLaguerre}, which facilitates the large-$N$ limit to be taken in Subsection \ref{lim}.

\subsection{The eigenvalue model and definition of its correlation functions}\label{Defs}

We begin by stating the RMT that we want to solve in its eigenvalue representation,  defined by the following partition function
\begin{equation}
Z_N^{(\beta,N_f,\nu)}=\left(\prod_{j=1}^N\int_0^\infty dx_j\ x_j^{\frac\beta2(\nu+1)-1}e^{-x_j}\prod_{f=1}^{N_f}(x_j+m_f^2)\right) |\Delta_N(\{x\})|^\beta= \left(\prod_{j=1}^N\int_0^\infty dx_j\right) {\cal P}_{jpdf}(x_1,\ldots,x_N)\ . 
\label{partitionfunct}
\end{equation}
It gives the normalisation constant of the unnormalised joint probability density function (jpdf) ${\cal P}_{jpdf}(x_1,\ldots,x_N)$ of all eigenvalues. Here we have included the dependence on the following parameters to be related to QCD later: $\nu=0,1,2,\ldots$ called the topological index taking fixed values, the masses $m_f\in\mathbf{R}_+$, $f=1,2,\ldots,N_f$, with $N_f$ counting the number of inserted  characteristic polynomials. In the quenched theory with $N_f=0$ their product is absent. These parameters  are all collected in the weight function
\begin{equation}
w(x)=x_j^{\frac\beta2(\nu+1)-1}e^{-x_j}\prod_{f=1}^{N_f}(x_j+m_f^2)\ ,
\label{weight}
\end{equation}
and finally we also have introduced the Vandermonde determinant
\begin{equation}
\Delta_N(\{x\})=\det_{1\leq i,j\leq N}[x_i^{j-1}]=\prod_{1\leq i<j\leq N}(x_j-x_i)\ .
\label{Delta}
\end{equation}
It depends on the set of all eigenvalues $\{x\}=\{x_i\}_{i=1,\ldots,N}$.
The partition function eq. (\ref{partitionfunct}) is relevant for QCD for $\beta=2$, representing the chGUE in the presence of $N_f$ characteristic polynomials (mass terms). For completeness and in order to state some open problems later we have also introduced the chGOE and chGSE with $\beta=1$ and 4, respectively. All three RMT are also called Wishart-Laguerre ensembles and can be written in matrix representation,
\begin{equation}
Z_N^{(\beta,N_f,\nu)}\sim\int[dW]\prod_{f=1}^{N_f}\det
\left[
\begin{array}{cc}
m_f 1_N&i W\\
iW^\dag &m_f 1_{N+\nu}\\
\end{array}
\right] \exp[-\mbox{Tr}(WW^\dag)]\ .
\label{matrixrep}
\end{equation}
Here $W$ is an $N\times(N+\nu)$ matrix taking values $W_{ij}\in\mathbf{R/C/H}$ for $\beta=1,2,4$ respectively, and $W^\dag$ is its Hermitian conjugate. The integration $[dW]$ is over the flat Lebesgue measure of all independent matrix elements. The identity matrices in front of the scalar mass variables $m_f$ of dimensions $N$ and $N+\nu$ are denoted by $1_N$ and $1_{N+\nu}$, respectively. 
Here we see that $\nu$ counts the number of zero eigenvalues of the Wishart matrix $W^\dag W$, and the $x_j$, $j=1,\ldots,N$ denote its non-zero positive eigenvalues\footnote{For quaternionic matrix elements the eigenvalues are doubly degenerate (Kramer's degeneracy), which is why for $\beta=4$ we always have an even number $2N_f$ of doubly degenerate masses in eq. (\ref{partitionfunct}). We will come back to this at the end of Subsection \ref{allLaguerre}.}. The change of variables from matrix elements to eigenvalues leads to the Vandermonde determinant to the power $\beta=1,2,4$ times the $\nu$-dependent part as the Jacobian, and we refer to standard text books on RMT for the derivation \cite{Mehta,BookPeter}.
In the lectures of J.-P. Bouchaud  these ensembles were discussed for $N_f=0$, in particular when $\nu$ is of the order of $N$. 
This matrix representation eq. (\ref{matrixrep}) will play an important role later in Section \ref{symm} when relating the RMT to QCD based on symmetries.
In the following we will mainly focus on the chGUE with $\beta=2$ which is the RMT relevant for QCD. The values $\beta=1$ and 4 apply to other QFTs and are very briefly discussed at the end of Subsection \ref{allLaguerre}.

What are the quantities we would like to calculate as functions of the masses $m_f$ and $\nu$, apart from the normalising partition function? The jpdf represents a determinantal (or Pfaffian) point process  introduced in the lectures by A. Borodin. Therefore we can determine all $k$-point eigenvalue density correlations defined for all $\beta$ as 
\begin{eqnarray}
R_k(x_1,\ldots,x_k)&=&\frac{1}{Z_N^{(\beta,N_f,\nu)}}\frac{N!}{(N-k)!}
\int_0^\infty dx_{k+1}\cdots\int_0^\infty dx_N {\cal P}_{jpdf}(x_1,\ldots,x_N)
\label{Rk}\\
&=&\prod_{j=1}^kw(x_j) \det_{1\leq i,j\leq N}[K_N(x_i,x_j)]\ ,\ \ \mbox{for} \ \ \beta=2\ .
\label{Rkb2}
\end{eqnarray}
In the second line valid for $\beta=2$ only they can be expressed in terms of the determinant of the kernel 
\begin{equation} 
K_N(x,y)= \sum_{l=0}^{N-1}h_l^{-1}P_l(x)P_l(y)\ ,
\label{kernel}
\end{equation}
that does not contain the weights in our convention. It contains 
OP with respect to the weight eq. (\ref{weight}), 
\begin{equation}
\int_0^\infty dx\ w(x) P_k(x)P_l(x)=h_k\delta_{kl}\ .
\label{OP}
\end{equation}
For $\beta=1,4$ similar expressions exist in terms of Pfaffian determinants of a matrix kernel of  OP with respect to a skew symmetric inner product replacing eq. (\ref{OP}). These are called skew OP and we refer to \cite{Mehta,BookPeter} for details.  
We choose the OP in eq. (\ref{OP}) to be monic, $P_k(x)=x^{k}+O(x^{k-1})$, with squared norms $h_k=||P_k||^2$. In particular eq. (\ref{Rk}) leads to the spectral density $R_1(x)=w(x)K_N(x,x)$ (normalised to $N$) and to the determinantal expression for the jpdf, $R_N(x_1,\ldots,x_N)=N!{\cal P}_{jpdf}(x_1,\ldots,x_N)/Z_N^{(\beta,N_f,\nu)}$.
Two further quantities can be defined that are of our interest. The $k$-th gap probability defined for all $\beta$ reads
\begin{equation}
E_k(s)=\frac{1}{Z_N^{(\beta,N_f,\nu)}}\frac{N!}{(N-k)!}
\int_0^sdx_1\cdots\int_0^sdx_k
\int_s^\infty dx_{k+1}\cdots\int_s^\infty dx_N {\cal P}_{jpdf}(x_1,\ldots,x_N)\ ,\ k=0,1,\ldots,N.
\label{gapk}
\end{equation}
It is the probability that $k$ eigenvalues are in $[0,s]$ and $N-k$ in $[s,\infty)$. For example $E_0(s)$ gives the gap probability that the interval $[0,s]$ is empty. If we order the eigenvalues $x_1<x_2<\ldots< x_N$ the $k$-th eigenvalue distribution defined as  
\begin{equation}
p_k(s)=\frac{k{{N}\choose{k}}}{Z_N^{(\beta,N_f,\nu)}}
\int_0^sdx_1\cdots\int_0^sdx_{k-1}
\int_s^\infty dx_{k+1}\cdots\int_s^\infty dx_N {\cal P}_{jpdf}(x_1,\ldots,x_k=s,\ldots,x_N),\ k=1,\ldots,N,
\label{pk}
\end{equation}
gives the probability to find the $k$-th eigenvalue at $s=x_k$, $k-1$ eigenvalues in $[0,s]$ and $N-k$ in $[s,\infty)$. These individual eigenvalue distributions are normalised $\int_0^\infty ds\,p_k(s)=1$, as can be easily checked (see e.g. the appendix of \cite{AD03}). We will be mostly interested in the example of the distribution of the smallest non-zero eigenvalue, $p_1(s)$.
Equations (\ref{gapk}) and (\ref{pk}) are not independent, in fact the latter follow by differentiation:
\begin{equation}
\frac{\partial}{\partial s} E_0(s)=-p_1(s)\ ,\ \ \frac{\partial}{\partial s} E_k(s)=k!(p_k(s)-p_{k+1}(s))\ ,\ k=1,\ldots,N-1, \ \frac{\partial}{\partial s} E_N(s)=N! p_N(s)\ ,
\label{p1}
\end{equation}
which can be solved for $p_k(s)$ as follows: 
\begin{equation}
p_k(s)=-\sum_{l=0}^{k-1}\frac{1}{l!}\frac{\partial}{\partial s} E_l(s)\ .
\end{equation}
The gap probabilities can also be expressed in terms of the kernel $K_N(x,y)$ as a Fredholm determinant, cf. the lectures by A. Borodin and eqs. (\ref{E0Andreiev}) and (\ref{expand}), but we will use a different route to compute them explicitly.

Our goal is now to determine all the quantities that we have introduced in this subsection, including the partition function. For $N_f=0$ this would be an easy task as the polynomials orthogonal with respect to $w(x)=x^\nu\exp[-x]$ are the well known Laguerre polynomials. It turns out that also for $N_f>0$ we can express all quantities in terms of Laguerre polynomials, including the dependence on the parameter $m_f$ and $\nu$  that we are after.  This will be the goal of the following 2 subsections.

\subsection{Properties of orthogonal polynomials with general weights}\label{genweight}

The first part of this subsection is standard material about OP on the real line and can be found in any standard reference, e.g. in \cite{Walter} including also OP in the complex plane needed later. For the second part with weights containing characteristic polynomials we refer to \cite{BDS} for more details.

Let us consider a general measurable weight function on the positive real half line with the condition that all moments exist, $M_k=\int_0^\infty dxw(x)x^k<\infty$. The monic polynomials $P_k(x)$ orthogonal with respect to $w(x)$ can then be recursively constructed using Gram-Schmidt, expressing them as the ratio of two determinants:
\begin{equation}
P_k(x)=
\left|
\begin{array}{llll}
M_0&M_1&\ldots&M_k\\
\vdots& \vdots& & \vdots\\
M_{k-1}&M_k&\ldots&M_{2k-1}\\
1&x&\ldots&x^k\\
\end{array}
\right|\left( \det_{0\leq i,j\leq k-1}[M_{i+j}]\right)^{-1},
\end{equation}
see \cite{Walter} for details. It obviously satisfies $P_k(x)=x^k+{\cal O}(x^{k-1})$.
These polynomials obey a three-step recurrence relation,
\begin{equation}
xP_k(x)=P_{k+1}(x)+\alpha_k^kP_k(x)+\alpha_k^{k-1}P_{k-1}(x)\ ,
\label{recurrence}
\end{equation}
which can be see as follows. With the $P_k(x)$ forming a complete set we can expand
\begin{equation}
xP_k(x)=\sum_{l=0}^{k+1}\alpha_k^l P_l(x)\ ,\ \ \alpha_k^l=h_l^{-1}
\int_0^\infty dxw(x)xP_k(x)P_l(x)\ .
\end{equation}
Using that in the last integral $xP_l(x)$ is a polynomial of degree $l+1$ and the orthogonality eq. (\ref{OP}) it follows that $\alpha_k^l=0$ for $l<k-1$ and 
$\alpha_k^{k-1}=h_k/h_{k-1}$. For the orthonormal polynomials $\hat{P}_k(x)=P_k(x)/\sqrt{h_k}$ the recursion eq. (\ref{recurrence}) becomes more symmetric, 
\begin{equation}
x\hat{P}_k(x)=c_k\hat{P}_{k+1}(x)+\alpha_k^k\hat{P}_k(x)+c_{k-1}\hat{P}_{k-1}(x)\ ,\ \ c_{k-1}=\sqrt{{h_{k}}/{h_{k-1}}}\ ,\ \ k\geq1\ ,
\label{recurrence2}
\end{equation}
and we obtain the Christoffel-Darboux formula for the kernel
\begin{equation}
K_N(x,y)= \sum_{l=0}^{N-1}\hat{P}_k(x)\hat{P}_k(y)=c_{N-1}
\frac{\hat{P}_N(x)\hat{P}_{N-1}(y)-\hat{P}_{N-1}(x)\hat{P}_N(y)}{x-y}\ ,\ x\neq y\ .
\label{CD}
\end{equation}
This simply follows by multiplying the sum by $(x-y)$ with $x\neq y$, using the  recursion and the fact that this is a telescopic sum. From l'H\^opital's rule for the kernel at equal arguments we see that in the asymptotic limit $N\to\infty$ we only need to evaluate the asymptotic of $P_N(x)$, $P_{N-1}(x)$, their derivatives, and the ratio of norms $c_N$ to determine the spectral density:
\begin{equation}
R_1(x)=w(x)K_N(x,x)= 
c_{N-1}({\hat{P}_N(x)\hat{P}_{N-1}^\prime(x)-\hat{P}_{N-1}(x)\hat{P}_N^\prime(x)}) .
\label{densityCD}
\end{equation}
Let us give two examples. For the GUE with weight $w(x)=\exp[-x^2/2]$ on $\mathbf{R}$ instead of $\mathbf{R}_+$ we have 
\begin{equation}
P_n(x)=\mbox{He}_n(x)\ ,\ \ h_n=\sqrt{2\pi}\,n!\ ,
\label{Hermite}
\end{equation}
the probabilist's Hermite polynomials. Second,
for the Laguerre weight $w(x)=x^\nu\exp[-x]$, eq. (\ref{weight}) with $N_f=0$ on 
$\mathbf{R}_+$, 
the monic OP are given by the appropriately rescaled generalised Laguerre polynomials $L_n^\nu(x)$:
\begin{equation}
P_n(x)=(-1)^nn!L_n^\nu(x)\ ,\ \ h_n=n!\ \Gamma(n+\nu+1)\ ,\ \ \nu>-1\ ,
\label{Laguerre}
\end{equation}
where $\nu$ can also be real here.
The asymptotic analysis of the kernel (and density) can now be made, and we will come back to this in Subsection \ref{lim}.

In the following we will express the partition function, corresponding OPs, kernel and gap probability $E_0(s)$ in terms of expectation values of characteristic polynomials in a first step, and then reduce them to determinants of Laguerre polynomials in a second step. In order to prepare this the following exercise is very useful.

Consider the 2 Vandermode determinants eq. (\ref{Delta}) inside the partition function (\ref{partitionfunct}) for $\beta=2$. By adding columns inside the Vandermonde determinant we do not change its value. Starting with the last column of highest degree we can thus express $\Delta_N(\{x\})$ in terms of arbitrary monic polynomials ${Q}_k(x)$:
\begin{equation}
\Delta_N(\{x\})=\det_{1\leq i,j\leq N}[{Q}_{j-1}(x_i)] \ .
\label{Delta-id1}
\end{equation}
Choosing those monic polynomials $Q_k(x)=P_k(x)$ that precisely satisfy the orthogonality relation (\ref{OP}) we obtain for the partition function:
\begin{eqnarray}
Z_N^{(2,N_f,\nu)}&=&\left(\prod_{l=1}^N\int_0^\infty dx_l w(x_l)\right)
\left(\det_{1\leq i,j\leq N}[{P}_{j-1}(x_i)]\right)^2\nonumber\\
&=&\sum_{\sigma,\sigma^\prime\in S_N}(-1)^{\sigma+\sigma^\prime}
\prod_{l=1}^N\int_0^\infty dx_l w(x_l){P}_{\sigma(l)-1}(x_l){P}_{\sigma^\prime(l)-1}(x_l)
\ =\ N!\prod_{l=1}^Nh_{l-1}\ .
\label{Znorm}
\end{eqnarray}
In the second step we have Laplace expanded both determinants into sums over permutations $\sigma,\sigma^\prime\in S_N$ and used the orthogonality (\ref{OP}). The computation of the partition function thus amounts to determine the squared norms of the OPs with respect to the weight eq. (\ref{weight}).

Let us define the expectation value of an operator ${\cal O}(\{x\})$ that only depends on the set of eigenvalues $\{x\}$:
\begin{equation}
\langle {\cal O}(\{x\})\rangle_N=\frac{1}{Z_N^{(\beta,N_f,\nu)}}
\left(\prod_{l=1}^N\int_0^\infty dx_{l}\right) {\cal P}_{jpdf}(x_1,\ldots,x_N)
{\cal O}(\{x\})\ . 
\label{ave}
\end{equation}
Then the following identity holds for the monic OPs:
\begin{equation}
P_L(x)=\left\langle \prod_{j=1}^L(x-x_j)\right\rangle_L.
\label{OPid}
\end{equation}
It is given by the expectation value of a characteristic polynomial with respect to $L$ eigenvalues for an arbitrary degree, where $0\leq L\leq N$. It can also be expressed in terms of matrices of size $L$, choosing  ${\cal O}(\{x\})=\det[x-WW^\dag]$. This relation is known as the Heine formula, dating back to the 19th century. The proof uses the simple identity for the Vandermonde determinant,
\begin{equation}
\prod_{j=1}^L(y-x_j)\Delta_L(\{x\})=
\Delta_{L+1}(\{x\},y=x_{L+1})=
\det_{1\leq i,j\leq L+1}[{P}_{j-1}(x_i)] \ .
\label{Delta-id2}
\end{equation}
Using this identity in the expectation value we arrive at the same situation as in eq. (\ref{Znorm}), with one larger determinant: 
\begin{eqnarray}
\left\langle \prod_{j=1}^L(y-x_j)\right\rangle_L&=& \frac{1}{Z_L^{(2,N_f,\nu)}}
\sum_{\stackrel
{\mbox{\small$\sigma\in S_L$}}{\mbox{\small$\sigma^\prime\in S_{L+1}$}}}
(-1)^{\sigma+\sigma^\prime}
\left(\prod_{l=1}^L\int_0^\infty dx_l w(x_l){P}_{\sigma(l)-1}(x_l){P}_{\sigma^\prime(l)-1}(x_l)\right)
{P}_{\sigma^\prime(L+1)-1}(y)
\nonumber\\
&=&P_L(y)\ .
\label{proofPN}
\end{eqnarray}
Here the orthogonality of the polynomials enforces $\sigma^\prime(L+1)=L+1$, and the remaining norms cancel due to eq. (\ref{Znorm}).
As a check also the leading coefficient of eq. (\ref{OPid}) can be compared, taking the asymptotic limit $x\gg1$ which leads to $x^{L}+{\cal O}(x^{L-1})$ on both sides.

Note that eq. (\ref{OPid}) provides an $L$-fold integral representation of the 
$P_L(x)$ for an arbitrary weight function. Compared to the known single integral representation, e.g. for the Hermite polynomials in the example in eq. (\ref{Hermite}), we thus have the following duality relation
\begin{equation}
\mbox{He}_n(x)=\frac{1}{\sqrt{2\pi}}\int_{-\infty}^\infty dt (x+it)^n\exp[-t^2/2]\ .
\end{equation}
Here 
we take a single average over $n$ copies of characteristic polynomials. The existence of such relations between 1- and $n$-fold averages of characteristic polynomials are known only for Gaussian weights. They can also be derived using the supersymmetric method as reviewed in chapter 7 of \cite{Handbook} by Thomas Guhr, see also \cite{MehtaNormand} for further dualities  using OP techniques.

In the next step we will express the kernel itself as an expectation value of two characteristic polynomials, following the work of P. Zinn-Justin \cite{PZJ}. Namely it holds that
\begin{equation}
K_{N+1}(x,y)=h_N^{-1}\left\langle \prod_{j=1}^N(x-x_j)(y-x_j)\right\rangle_N .
\label{Kernelid}
\end{equation}
The proof uses the same idea as before, now including each product into a different Vandermonde determinant of size $N+1$. We have 
\begin{eqnarray}
\left\langle \prod_{j=1}^N(x-x_j)(y-x_j)\right\rangle_N&=&
 \frac{1}{Z_N^{(2,N_f,\nu)}}
\sum_{\sigma,\sigma^\prime\in S_{N+1}}
(-1)^{\sigma+\sigma^\prime}
\left(\prod_{l=1}^N\int_0^\infty dx_l w(x_l){P}_{\sigma(l)-1}(x_l){P}_{\sigma^\prime(l)-1}(x_l)\right)\nonumber\\
&&\times{P}_{\sigma(N+1)-1}(x){P}_{\sigma^\prime(N+1)-1}(y)\nonumber\\
&=&
 \frac{1}{Z_N^{(2,N_f\nu)}}N!
\sum_{\sigma(N+1)=\sigma^\prime(N+1)=1}^{N+1}\prod_{j=1}^{N+1}h_{j-1}
\frac{1}{h_{\sigma(N+1)-1}}{P}_{\sigma(N+1)-1}(x){P}_{\sigma^\prime(N+1)-1}(y)
\nonumber\\
&=& \frac{1}{Z_N^{(2,N_f,\nu)}}N!\prod_{j=1}^{N+1}h_{j-1}
\sum_{l=0}^N \frac{P_l(x)P_l(y)}{h_l}
=\frac{P_{N+1}(x)P_N(y)-P_N(x)P_{N+1}(y)}{x-y}.\ \ \ \ \ \ 
\label{proofKernel}
\end{eqnarray}
Compared to eq. (\ref{proofPN}) the orthogonality only implies $\sigma(N+1)=\sigma^\prime(N+1)$, and its value can still run over all possible values from 1 to $N+1$. In the last line we give the expression for the kernel $K_{N+1}(x,y)$ in terms of the Christoffel-Darboux formula (\ref{CD}), using monic polynomials instead. Once again the leading coefficients of left and right hand side can be see to agree $\sim (xy)^N$ in the limit $x,y\gg1$.

An immediate question arises: What is the expectation value for more than two products of characteristic polynomials? For that purpose let us introduce some more notation following Baik, Deift and Strahov \cite{BDS}, to where we refer for more details. Denote by $P_{n}^{[l]}(x)$ the OP with respect to weight $w^{[l]}(x)=(\prod_{j=1}^l(y_j-x))w(x)$ for $l=1,2,\ldots$, with $P_{n}^{[0]}(x)=P_n(x)$. The following relation holds which is called Christoffel formula:
\begin{equation}
P_{n}^{[l]}(x)=\frac{1}{(x-y_1)\cdots(x-y_l)}
\left|
\begin{array}{lll}
P_n(y_1)&\ldots&P_{n+l}(y_1)\\
\vdots& & \vdots\\
P_n(y_l)&\ldots&P_{n+l}(y_l)\\
P_n(x)&\ldots&P_{n+l}(x)\\
\end{array}
\right|\left(\det_{1\leq i,j\leq l}[P_{n+j-1}(y_i)]\right)^{-1}.
\label{Christoffel}
\end{equation}
As explained in \cite{BDS} this can be seen as follows. 
It is clear that the determinant in the middle,  which we call $q_n^{[l]}(x)$, is a polynomial of degree $n+l$ in $x$. It has zeros  at $x=y_1,\ldots,y_l$, and thus 
$q_n^{[l]}(x)/(x-y_1)\cdots(x-y_l)$ is a polynomial of degree $n$ in $x$.  Thus we have that 
\begin{equation}
\int_0^\infty dx\, x^j\frac{q_n^{[l]}(x)}{(x-y_1)\cdots(x-y_l)}w^{[l]}(x)=0\ , \mbox{for}\ \ j=0,1,\ldots,n-1\ ,
\end{equation}
due to $P_n(x)$ being OP with respect to weight $w(x)$, after cancelling the factors from $w^{[l]}(x)$. The last factor in 
eq. (\ref{Christoffel}) ensures that  $P_{n}^{[l]}(x)$ is monic, as can bee seen from taking the limit $x\gg1$.
The repeated product of the Christoffel formula (\ref{Christoffel}) leads to the following theorem (see Theorem 2.3 \cite{BDS}) for the average of products of $l+1$ characteristic polynomials with $l=1,2,\ldots$, which is due to Br\'ezin and Hikami \cite{BH}:
\begin{equation}
\frac{1}{\Delta_{l+1}(\{y\})}\det_{1\leq i,j\leq l+1}[P_{n+j-1}(y_i)]
=\prod_{j=0}^lP_n^{[j]}(y_{j+1})=
\left\langle 
\prod_{i=1}^{n}\left(\prod_{j=1}^{l+1}(y_j-x_i) \right)\right\rangle_n.
\label{prodchar}
\end{equation}
For the last equality we have used eq. (\ref{OPid}) that leads to
\begin{equation}
P_n^{[j]}(y_{j+1})=\frac{\left\langle 
\prod_{i=1}^{n}\left((y_{j+1}-x_i)\prod_{p=1}^{j}(y_p-x_i) \right)\right\rangle_n}{\left\langle 
\prod_{i=1}^{n}\left(\prod_{p=1}^{j}(y_p-x_i) \right)\right\rangle_n}.
\label{Pnvev}
\end{equation}
In the product almost all factors cancel out. The case for a single characteristic polynomial with $l+1=1$ in eq. (\ref{prodchar}) was already stated eq. (\ref{OPid}).

In \cite{BDS} in Theorem 3.2 a further identity was derived for the expectation value of the product of an even number of characteristic polynomials, expressing it through a determinant of kernels divided by two Vandermonde determinants,
\begin{equation}
\frac{\prod_{l=N}^{N+K-1}h_l}{\Delta_{K}(\{\lambda\})\Delta_K(\{\mu\})}
\det_{1\leq i,j\leq K}[K_{N+K}(\lambda_i,\mu_j)]=
\left\langle 
\prod_{i=1}^{N}\left(\prod_{j=1}^{K}(\lambda_j-x_i) (\mu_j-x_i)\right)\right\rangle_N.
\label{prod2char}
\end{equation}
The simplest example for this relation was derived in eq. (\ref{Kernelid}). In \cite{AV03} this set of identities  was further generalised, expressing the expectation value of arbitrary products of characteristic polynomials by a determinant containing both kernels and polynomials divided by two Vandermonde determinants in many different and equivalent ways, cf. eq. (\ref{Cprodchar}). The simplest example with 3 products reads \cite{AV03}
\begin{equation}
\frac{h_N}{v_2-v_1}
\left|\begin{array}{ll}
K_{N+1}(v_1,u)&P_{N+1}(v_1)\\
K_{N+1}(v_2,u)&P_{N+1}(v_2)\\
\end{array}\right|=
\left\langle \prod_{i=1}^{n}(v_1-x_i)(v_2-x_i)(u-x_i)\right\rangle_N.
\end{equation}

Let us mention that in \cite{BDS} also expectation values of ratios of characteristic polynomials were determined. They can be expressed in terms of the Cauchy transforms $C_k(x)$ of the monic polynomials $P_k(x)$,
\begin{equation}
{C}_k(y)=\frac{1}{2\pi i}\int dt \frac{P_k(t)}{t-y}w(t)\ ,\ y\in\mathbf{C}\setminus\mathbf{R}\ ,
\label{Cauchy}
\end{equation}
as well as through mixed Chistoffel-Darboux kernels containing both Cauchy transforms and polynomials. The simplest example is given by 
\begin{equation}
C_{L-1}(x)=\frac{-h_{L-1}}{2\pi i}\left\langle \frac{1}{\prod_{j=1}^L(x-x_j)}\right\rangle_L.
\label{Cauchyid}
\end{equation}
The fact that both sides agree $\sim x^{-L}$ for $x\gg1$ can be easily seen from the definition (\ref{Cauchy}), expanding the geometric series and using the orthogonality of the polynomial $P_k(t)$ to monic powers less than $k$.

A second particularly important example is the expectation value of the ratio of two characteristic polynomials, due to the following relation to the resolvent or Stieltjes transform $G(x)$:
\begin{equation}
G_N(x)=\left\langle \sum_{j=1}^N\frac{1}{x-x_j}\right\rangle_N
=\frac{\partial}{\partial x}\left.\left\langle \frac{\prod_{j=1}^N(x-x_j)}{\prod_{j=1}^N(y-x_j)}\right\rangle_N\right|_{x=y}.
\label{resolvent}
\end{equation}
The resolvent can be used to obtain the spectral density through the following relation
\begin{equation}
R_1(x)=\frac{-1}{2\pi i}\lim_{\epsilon\to 0^+}[G_N(x+i\epsilon)-G_N(x-i\epsilon)] \ .
\label{inversion}
\end{equation}
There are many examples where the OP technique is not available, but where the expectation value of ratios of characteristic polynomials can be found by other means, e.g. by using supersymmetry, replicas or loop equations. This then leads to an alternative way to determine the spectral density, or higher $k$-point correlation functions, by taking the average over $k$ resolvents and then taking the imaginary parts with respect to each of the $k$ arguments. We refer to chapters 7, 8 and 16 in \cite{Handbook}  for further details and references.

\subsection{All correlation functions with masses in terms of Laguerre polynomials}\label{allLaguerre}

Using the results from the previous subsection we are now ready to express the partition function, OPs, kernel, and consequently all $k$-point correlation functions with $N_f$ characteristic polynomials, as well as the corresponding gap probabilities through expectation values of characteristic polynomials with respect to the quenched weight $w(x)=x^\nu\exp[-x]$ with $N_f=0$. Because the OPs of this quenched weight are Laguerre polynomials, everything will be finally expressed through these, which yields explicit and exact expressions for any finite $N$ and $N_f$. In the next Subsection \ref{lim} will use them to take the large-$N$ limit based on the known asymptotic of the Laguerre polynomials.

We shall adopt the notation from the previous section, labelling all the above listed quantities by superscript $[N_f]$ compared to the quenched weight without superscript. We have for the partition function eq. (\ref{partitionfunct}) with masses
\begin{eqnarray}
Z_N^{[N_f]}(\{m\})&=&\left(\prod_{j=1}^N\int_0^\infty dx_j\ x_j^\nu e^{-x_j}\prod_{f=1}^{N_f}(x_j+m_f^2)\right)\Delta_N(\{x\})^2
=Z_N\left\langle \prod_{j=1}^N\prod_{f=1}^{N_f}(x_j+m_f^2)\right\rangle_N
\nonumber\\
&=&N!(-1)^{N_f(N+(N_f-1)/2)}\prod_{j=1}^{N}\Gamma(j+\nu)\prod_{j=1}^{N+N_f}\Gamma(j)\ 
\frac{\det_{1\leq i,j \leq N_f}
\left[L_{N+j-1}^\nu(-m_i^2)\right]}{\Delta_{N_f}(\{-m^2\})}.
\label{Zmassive}
\end{eqnarray}
This result simply follows from choosing $y_j=-m_j^2$ in eq. (\ref{prodchar}), together with eqs. (\ref{Znorm}) and (\ref{Laguerre}) for the quenched polynomials. It is clear that if some of the masses become zero, say $L$ out of $N_f$, from looking at eq. (\ref{partitionfunct}) for $\beta=2$ without further calculation this leads to the shift $\nu\to\nu+L$ in the remaining determinant of size $N_f-L$. This property is called flavour-topology duality. In the case when all masses are degenerate, $m_f=m\ \forall f$, l'Hopital's rule eventually leads to \cite{BH}
\begin{equation}
Z_N^{[N_f]}(m)=\frac{(-1)^{NN_f}}{\prod_{l=0}^{N_h-1}l!}\det_{0\leq i,j \leq N_f-1} \left[(-1)^{N+i}(N+i)!(L_{N+i}^\nu(-m^2))^{(j)}\right]\ ,
\label{Zmassivedeg}
\end{equation}
where the superscript $^{(j)}$ denotes the $j$-th derivative with respect to the argument $x=-m^2$.

The unquenched OP $P_n^{[N_f]}(x)$ follow directly from eq. (\ref{Pnvev}) at $j=N_f$, with $x=y_{j+1}$. For example using eq. (\ref{Laguerre}) the OP for the weight $w^{[1]}(x)=(x+m^2)x^\nu\exp[-x]$ the corresponding OP read \cite{DN}
\begin{equation}
P_n^{[N_f=1]}(x)=\frac{h_nK_{n+1}(x,-m^2)}{P_n(-m^2)}=\frac{(-1)^{n+1}(n+1)!}{(x+m^2)L_n^\nu(-m^2)}(L_{n+1}^\nu(x)L_{n}^\nu(-m^2)-L_{n+1}^\nu(-m^2)L_{n}^\nu(x))\ .
\label{Pmassive}
\end{equation}
However, it is much simpler to directly use the kernel from eq. (\ref{Kernelid}), rather than expressing it through the polynomials $P_N^{[N_f]}(x)$ using the Christoffel-Darboux identity (\ref{CD}). Indeed the kernel is given by eq. (\ref{Kernelid}) as
\begin{eqnarray}
K_N^{[N_f]}(x,y)&=& \frac{1}{h_{N-1}^{[N_f]}}\left\langle
\prod_{i=1}^{N-1}(x-x_i)(y-x_i)
\right\rangle_{N-1}^{[N_f]}
=\frac{1}{h_{N-1}}\frac{
\left\langle\prod_{i=1}^{N-1}\left((x-x_i)(y-x_i)\prod_{f=1}^{N_f}(m_f^2+x_i)\right)
\right\rangle_{N-1}
}{
\left\langle\prod_{i=1}^{N}\prod_{f=1}^{N_f}(m_f^2+x_i)
\right\rangle_{N}
}\nonumber\\
&=& \frac{(-1)^{N_f-1}(N+N_f)!}{\Gamma(N+\nu)(y-x)\prod_{f=1}^{N_f}(y+m_f^2)(x+m_f^2)}
\frac{
\left|
\begin{array}{lll}
L_{N-1}^\nu(-m_1^2)&\ldots&L_{N+N_f}^\nu(-m_1^2)\\
\vdots& & \vdots\\
L_{N-1}^\nu(-m_{N_f}^2)&\ldots&L_{N+N_f}^\nu(-m_{N_f}^2)\\
L_{N-1}^\nu(x)&\ldots&L_{N+N_f}^\nu(x)\\
L_{N-1}^\nu(y)&\ldots&L_{N+N_f}^\nu(y)\\
\end{array}
\right|
}{\det_{1\leq f,g\leq N_f}[L_{N+g-1}^\nu(-m_f^2)]}.
\label{Kernelmassive}
\end{eqnarray}
In the first step after inserting eq. (\ref{Kernelid}) we have included the mass dependent norm $h_{N-1}^{[N_f]}$ into the massive partition function in the denominator. It is given by the product of its norms times $N!$. Thus we have increasing the average from $N-1$ to $N$. In the final result in the second line we have already taken out all signs and factorials from the two determinants and cancelled them partly. Also the ratio of the two Vandermonde determinants, one of which containing the arguments $x$ and $y$ has been simplified using the definition (\ref{Delta}). The kernel eq. (\ref{Kernelmassive}) determines all $k$-point eigenvalue correlation functions from eq. (\ref{Rkb2}) including their mass dependence. In particular for $x=y$ we obtain the spectral density 
with $N_f$ masses:
\begin{equation}
R_1^{[N_f]}(x)=x^\nu\exp[-x]\prod_{f=1}^{N_f}(x+m_f^2)\ K_N^{[N_f]}(x,x)\ ,
\label{R1massive}
\end{equation}
after applying l'H\^opital's rule once. This leads to $L^\nu_{N+g-1}(x)^\prime$ in the last row in the numerator. Note that the product from the weight in front of the kernel cancels part of its denominator.

Let us now turn to the gap probabilities and distribution of smallest eigenvalues, starting with $E_0(s)$ from eq. (\ref{gapk}).
The following Andr\'eiev integral identity (cf. Borodin's lectures) can be used to get a first expression:
\begin{equation}
\int dx_1 \ldots\int dx_N 
\det_{1\leq i,j \leq N}[\phi_i(x_j)]\det_{1\leq i,j \leq N}[\psi_i(x_j)]
=N!\det_{1\leq i,j \leq N}\left[\int dx\phi_i(x)\psi_j(x)\right]\ .
\label{Andreiev}
\end{equation}
The only condition to hold is that all integrals of the functions $\phi_j(x)$ and $\psi_j(x)$ exist. The proof merely uses the Laplace expansion of the left hand side: 
\begin{equation}
\sum_{\sigma,\sigma^\prime\in S_N}(-1)^{\sigma+\sigma^\prime}\prod_{j=1}^N\int dx_{\sigma^\prime(j)} \phi_{\sigma^{-1}(\sigma^\prime(j))}(x_{\sigma^\prime(j)}) \psi_j(x_{\sigma^\prime(j)}) =N! \sum_{\sigma^{\prime\prime}\in S_N}(-1)^{\sigma^{\prime\prime}}\prod_{j=1}^N \int dx \phi_{\sigma^{\prime\prime}(j)}(x)\psi_j(x) \ ,
\label{proofAnd}
\end{equation}
and integration over common arguments of $\phi_i(x_{\sigma(i)}=x_{\sigma^\prime(j)})$ and $\psi_j(x_{\sigma^\prime(j)})$. This implies $i=\sigma^{-1}(\sigma^\prime(j))=\sigma^{\prime\prime}(j)$, which is yet another permutation.
Following  eq. (\ref{gapk}) we can thus write 
\begin{equation}
E_0^{[N_f]}(s) =\frac{1}{N!
}
\prod_{p=1}^N\int_s^\infty dx_p 
\det_{1\leq i,j \leq N}\left[\sqrt{w^{[N_f]}(x_j)}\hat{P}_{i-1}^{[N_f]}(x_j)\right]^2
=\det_{1\leq i,j \leq N}\left[\delta_{ij}-\int_0^s dx w^{[N_f]}(x)\hat{P}_{i-1}^{[N_f]}(x)\hat{P}_{j-1}^{[N_f]}(x)\right]\!,
\label{E0Andreiev}
\end{equation}
which can be interpreted as a Fredholm determinant.
In the first step we have replaced the Vandermonde determinants by determinants of monic polynomials, and then included the square root of the weight functions and of the norms stemming from the normalising partition function into the determinants, in order to make the polynomials orthonormal, $\hat{P}_{i-1}^{[N_f]}(x_j)$. In the next step we have used the Andr\'eief formula and the orthonormality on $[0,\infty)$. Knowing both the monic polynomials from eq. (\ref{Pnvev}) and their squared norms from $h_{j-1}^{[N_f]}=Z_j^{[N_f]}/(j Z_{j-1}^{[N_f]})$ in terms of expectation values of Laguerre polynomials, eq. (\ref{E0Andreiev}) is a valid expression for the gap $E_0(s)$. However, we will use a more direct expression to be derived below. As a further remark we note that replacing $\int_s^\infty dx =\int_0^\infty dx - \int_0^s dx$ in the first equality in  (\ref{E0Andreiev}) we obtain an equivalent expansion of the Fredholm determinant in the second equation:
\begin{equation}
E_0^{[N_f]}(s) =1+\sum_{k=1}^N\frac{(-1)^k}{k!}\int_0^s dx_1\ldots \int_0^s dx_k R_k^{[N_f]}(x_1\ldots,x_k).
\label{expand}
\end{equation}
In order to express the gap probability directly as an expectation value let us go back to its definition (\ref{gapk}), which we state for general $\beta$:
\begin{eqnarray}
E_0^{[N_f]}(s)&=&\frac{1}{Z_N^{(\beta,N_f,\nu)}}\left(\prod_{j=1}^N\int_s^\infty dx_j\ x_j^{\frac\beta2(\nu+1)-1}e^{-x_j}\prod_{f=1}^{N_f}(x_j+m_f^2)\right) |\Delta_N(\{x\})|^\beta\nonumber\\
&=&\frac{e^{-N s}}{Z_N^{(\beta,N_f,\nu)}}\left(\prod_{j=1}^N\int_0^\infty dy_j\ y_j^0 (y_j+s)^{\frac\beta2(\nu+1)-1}e^{-y_j}\prod_{f=1}^{N_f}(y_j+s+m_f^2)\right) |\Delta_N(\{y\})|^\beta.
\label{gap-shift}
\end{eqnarray}
Here we have shifted all integration domains, substituting $x_j=y_j+s$, $j=1,\ldots,N$. 
The shift can be simply worked out, in particular it leaves the Vandermonde determinant invariant,
As a consequence the numerator is again given by a partition of $N_f+\frac\beta2(\nu+1)-1$ masses with values $\sqrt{m_f^2+s}=m_f^\prime$ for the first $N_f$, and $\sqrt{s}$ for the remaining ones, with an effective topological charge $\nu_{ef\!f}$ parameter satisfying $\frac\beta2(\nu_{ef\!f}+1)-1=0$.
This idea has been used in several papers \cite{ForresterHughes,GWW,DNW,TaroPeter,DN} to give closed form expressions as an alternative to the Fredholm determinant above.

Let us first consider $\beta=2$, with $\frac\beta2(\nu+1)-1=\nu$. Consequently eq. (\ref{gap-shift}) can be written as 
\begin{equation}
\underline{\beta=2:}\ \ E_0^{[N_f]}(s)= e^{-Ns} \frac{Z_N^{(2,N_f+\nu,0)}}{Z_N^{(2,N_f,\nu)}}
\left\langle \prod_{i=1}^N\left( (s+y_j)^\nu\prod_{f=1}^{N_f}(m_f^2+s+y_j)\right)\right\rangle_{N,\,\nu_{ef\!f}=0}.
\label{gap-vevb2}
\end{equation}
The simplest examples with $N_f=0$ and $\nu=0,1$ thus read
\begin{eqnarray}
\nu=0:\ \ E_0(s)&=& e^{-Ns}\ ,
\label{E0nu0}\\
\nu=1:\ \ E_0(s)&=& e^{-Ns}L_N^0(-s){{N+\nu}\choose{N}}^{-1}.
\label{E0nu1}
\end{eqnarray}
In the second line we have replaced the expectation value of a single characteristic polynomial eq. (\ref{OPid}) by the Laguerre polynomial eq. (\ref{Laguerre}) and fixed the normalisation by the requirement $E_0(s=0)=1$. 
The quenched gap probability ($N_f=0$) with arbitrary $\nu$ can be computed from the massive partition functions eq. (\ref{Zmassivedeg}) at complete degeneracy.

The smallest eigenvalue distribution $p_1(s)$ easily follows for the examples we have just given by differentiation of the gap probability, see eq. (\ref{p1}). For higher gap probabilities and for computing $p_k(s)$ directly the same trick from eq. (\ref{gap-shift}) can be used. First consider only the integrals $\int_s^\infty dx$ in the definitions (\ref{gapk}) and (\ref{pk}). Then do the shift $x_j=y_j+s$, and perform the remaining  integrals  $\int_0^s dx$ over the obtained expectation value of characteristic polynomials in the end.
We refer to \cite{DN} for more details.

Let us now briefly comment on $\beta=1,4$, mainly because of recent developments and open questions in these symmetry classes. We begin with $\beta=1$. Here the effective topological charge is $\nu_{ef\!f}=1$ to satisfy  $0=(\nu_{ef\!f}-1)/2$ in eq. (\ref{gap-shift}). The 
additional mass terms originating from the shift by $s$ in  eq. (\ref{gap-shift}) appear with multiplicity $(\nu-1)/2$: 
\begin{equation}
\underline{\beta=1:}\ \ E_0^{[N_f]}(s)= e^{-Ns} \frac{Z_N^{(1,N_f+(\nu-1)/2,1)}}{Z_N^{(1,N_f,\nu)}}
\left\langle \prod_{i=1}^N\left( (s+y_j)^{(\nu-1)/2}\prod_{f=1}^{N_f}(m_f^2+s+y_j)\right)\right\rangle_{N,\,\nu_{ef\!f}=1}.
\label{gap-vevb1}
\end{equation}
That is for 
$\nu=2l+1$ odd, $l\in\mathbf{N}$ we have $l$ extra masses $\sqrt{s}$ compared to the $N_f$ shifted masses. Here the simplest case is $l=0$
with $N_f=0$, for which we obtain
\begin{equation}
\nu=1:\ \ E_0(s)= e^{-Ns}
\label{E0nu1b1}
\end{equation}
for arbitrary $N$. It agrees with the result for $\beta=2$ at $\nu=0$, eq. (\ref{E0nu0}).
For general odd $\nu$ the gap probabilities have been computed and we refer to \cite{DN} for results.
However,  for $\nu=2l$ even we obtain a half integer number $l-1/2$ of additional mass terms. This leads to the question of calculating expectation values including products of square roots of characteristic polynomials in eq. (\ref{gap-vevb1}). 
In \cite{AGKWW} the expectation values needed to determine $E_0(s)$ for arbitrary even $\nu=2l$ and general $N_f$ have been computed for the chGOE, answering at least the question for the gap probability. The cases $\nu=0$ \cite{Peter93} and $\nu=2$ \cite{AVivo} were previously known from different considerations. 
Namely in \cite{Edelman} a recursive construction was made for the smallest eigenvalue for $\beta=1$ valid for all $\nu$. However, in this approach the Pfaffian structure appearing for $\nu\geq4$ is not at all apparent.

Independently in \cite{FNock} the question about expectation values of square roots of determinants has been asked and answered for special cases for the GOE in the context of scattering in chaotic quantum systems. It is an open problem if the most general expectation value of such products (or ratios) of square roots has a 
determinantal or Pfaffian structure as in eq. (\ref{prodchar}).

Let us turn to $\beta=4$. The effective topological charge we obtain is $\nu_{ef\!f}=-1/2$ to satisfy  $0=2\nu_{ef\!f}+1$ in eq. (\ref{gap-shift}).
The fact that it is not integer does not pose a problem as the index of the generalised Laguerre polynomials eq. (\ref{Laguerre}) in terms of which we have expressed expectation values of characteristic polynomials can be chosen accordingly. The more sever problem is the following. Due to Kramer's degeneracy the eigenvalues $x_j$ of a self-dual quaternion valued $N\times N$ matrix $H\in\mathbf{H}$ always come in pairs, implying $\det[x-H]=\prod_{j=1}^N(x-x_j)^2$. Consequently the mass terms generated from eq. (\ref{matrixrep}) always occur with an {\it even} power $N_f$ of two-fold degenerate masses in eq. (\ref{partitionfunct}). In contrast in eq. (\ref{gap-shift}) we always need to evaluate an odd number of powers $(y_j+s)^{2\nu+3}$. Thus as for $\beta=1$ with even $\nu$ we need to evaluate square roots of determinants in order to compute the gap probability for $\beta=4$. This is an open problem so far and only the gap probability (and smallest eigenvalue distribution) with $\nu=0$ is known explicitly \cite{Peter93}. A Taylor series expansion exists though for $\nu>0$, following from group integrals of Kaneko-type \cite{Tiloetal}.

\subsection{The large-$N$ limit at the hard edge}\label{lim}

After having exhaustively  presented the solution of the eigenvalues model eq. (\ref{partitionfunct}) and its correlation functions for a finite number of eigenvalues $N$ at arbitrary $N_f$ for $\beta=2$, we will now turn to the large-$N$ limit. As it is true in general in RMT one has to distinguish between different large-$N$ scaling regimes. The global spectral statistics, which includes the global macroscopic density  given by the semi-circle law for the GUE, or the Marchenko-Pastur law for the chGUE, see Figure \ref{fig:MPdensity}, is concerned with correlations between eigenvalues that have many other eigenvalues (in fact a finite fraction of all) in between them. For a discussion of this large-$N$ limit alternative techniques are available such as loop-equations, and we refer to \cite{Amb} for a standard reference where all correlation functions including subleading contributions are computed recursively for $\beta=2$. In this limit the fluctuations of the eigenvalues on a local scale (of a few eigenvalues) are averaged out, and typically expectation values factorise.  

In contrast when magnifying the fluctuations among eigenvalues at a distance of $1/N^\delta$ one speaks of microscopic limits. The value of $\delta$ and the form of the limiting kernel depend on the location in the spectrum, see Figure \ref{fig:MPdensity}. At the so-called soft edge of the Marchenko-Pastur law one finds the Airy-kernel, whereas in the bulk of the spectrum the sine-kernel is found, see e.g. \cite{BookPeter}. Here we will be interested in the eigenvalues in the vicinity of the origin presenting a hard edge with $\delta=1/2$ and the limiting kernel to be computed below is the Bessel-kernel. Why this limit is relevant for the application of RMT to QCD has been indicated already in the introduction after eq. (\ref{BCrel}).

As a further consequence of this comparison to QCD we will change variables from Wishart eigenvalues $x_j$ of $WW^\dag$ to Dirac operator eigenvalues $y_j=\pm\sqrt{x_j}$ (more details follow in Section \ref{symm}). Looking at the partition function eq. (\ref{partitionfunct}) the Dirac eigenvalues and masses have to be rescaled with the same power in $N$. Otherwise we would immediately loose the dependence on the masses. On the RMT side the hard-edge scaling limit zooming into the vicinity of the origin is conventionally defined including factors by taking $N\to\infty$ and $y_j,m_f\to0$, such that the following product remains finite\footnote{Note that we consider weights that are $N$-independent here in contrast to other conventions.}
\begin{equation}
\tilde{y}_j=\lim_{\stackrel{\mbox{\small $N\to\infty$}}{\mbox{\small $y_j\to0$}}} 2\sqrt{N}\,y_j\ ,\ \ 
\tilde{m}_f=\lim_{\stackrel{\mbox{\small $N\to\infty$}}{\mbox{\small $m_f\to0$}}} 2\sqrt{N}\,m_f\ .
\label{RMTmass}
\end{equation}
In the last subsection we have eventually expressed all quantities of interest in terms of the Laguerre polynomials of the quenched weight. Therefore the only piece of information we need to take the asymptotic limit of all these quantities is the well known limit \cite{GradshteynRyzhik}:
\begin{equation}
\lim_{N\to\infty} N^{-\nu} L_N^\nu\left( \frac{x}{N}\right)=x^{-\nu/2}J_\nu\left( 2\sqrt{x}\right) \ .
\label{LaguerreLim}
\end{equation}
Consequently all Laguerre polynomials of positive argument turn into $J$-Bessel functions. Those polynomials containing negative arguments will turn into $I$-Bessel functions, due to the relation $I_\nu(z)=i^{-\nu} J_\nu(iz)$ for integer $\nu$. Note that both types of Bessel functions also appear together, see e.g. in eqs. (\ref{Pmassive}) and (\ref{Kernelmassive}). 
The asymptotics of the Laguerre OP eq. (\ref{LaguerreLim}) in fact holds for OP with respect to much more general weight functions. This phenomenon is called universality. After first results in \cite{ADMN,DN98,DNW} more sophisticated rigorous mathematical methods including the Riemann-Hilbert approach were developed, see chapter 6 by A. Kuijlaars in \cite{Handbook} for a detailed discussion and for references.

In the following we will give a few examples for the hard edge limit, starting with the partition function. Looking at eq. (\ref{Zmassive}) it is clear that we have to normalise it differently, such that the limit exists and gives a function of the limiting masses from eq. (\ref{RMTmass}). For $N_f=1$ this is very easy to do by dividing out the quenched partition function, and we define:
\begin{equation}
{\cal Z}_N^{[N_f=1]}(m)=m^{\nu}N^{-\frac{\nu}{2}}\frac{Z_N^{[N_f=1]}(m)}{Z_N N!}= m^{\nu}N^{-\frac{\nu}{2}} L_N^\nu(-m^2)\ ,
\label{ZRMTredef}
\end{equation}
leading to the limit
\begin{equation}
\lim_{\stackrel{\mbox{\small $N\to\infty$}}{\mbox{\small $m\to0$}}} {\cal Z}_N^{[N_f=1]}(m)={\cal Z}^{[N_f=1]}(\tilde{m})= I_\nu(\tilde{m})\ .
\label{ZRMTNf1}
\end{equation}
Note that our conventions are chosen such that for small mass the limiting partition function behaves as ${\cal Z}_N^{[N_f=1]}(\tilde{m}\approx0)\sim \tilde{m}^\nu$, rather than unity. The reason for this choice will also become transparent in the next section. 
For general $N_f$ we do not spell out the normalisation constant explicitly and directly give the limiting result for non-degenerate masses 
\begin{equation}
\lim_{\stackrel{\mbox{\small $N\to\infty$}}{\mbox{\small $m_f\to0$}}} {\cal Z}_N^{[N_f]}(\{m\})={\cal Z}^{[N_f]}(\{\tilde{m}\})= 
2^{\frac{N_f(N_f-1)}{2}}\prod_{j=0}^{N_f-1}\Gamma(j+1)\ 
\frac{\det_{1\leq f,g\leq N_f}\left[\tilde{m}_f^{g-1}I_{\nu+g-1}(\tilde{m}_f)\right]}{\Delta_{N_f}(\{\tilde{m}^2\})}\ ,
\label{ZRMTNf}
\end{equation}
where we adopted the normalisation from \cite{Baha}.  In the completely degenerate limit this leads to:
\begin{equation}
{\cal Z}^{[N_f]}(\tilde{m})
= {\det_{1\leq f,g\leq N_f}\left[I_{\nu+g-f}(\tilde{m})\right]}\ .
\label{ZRMTNfdeg}
\end{equation}
The direct limit $N\to\infty$ of eq. (\ref{Zmassivedeg}) (or l'Hopital's) rule from eq. (\ref{ZRMTNf})) leading to a determinant of derivatives of $I$-Bessel functions can be brought to this form of a Toeplitz determinant using identities for Bessel functions.

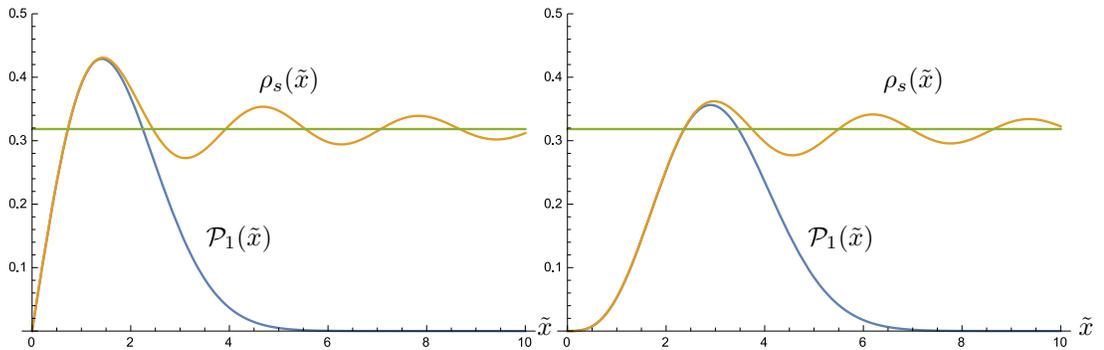
\begin{figure}[h]
\unitlength1cm
\begin{center}
\begin{picture}(12.2,5.5)
\includegraphics[width=7cm]{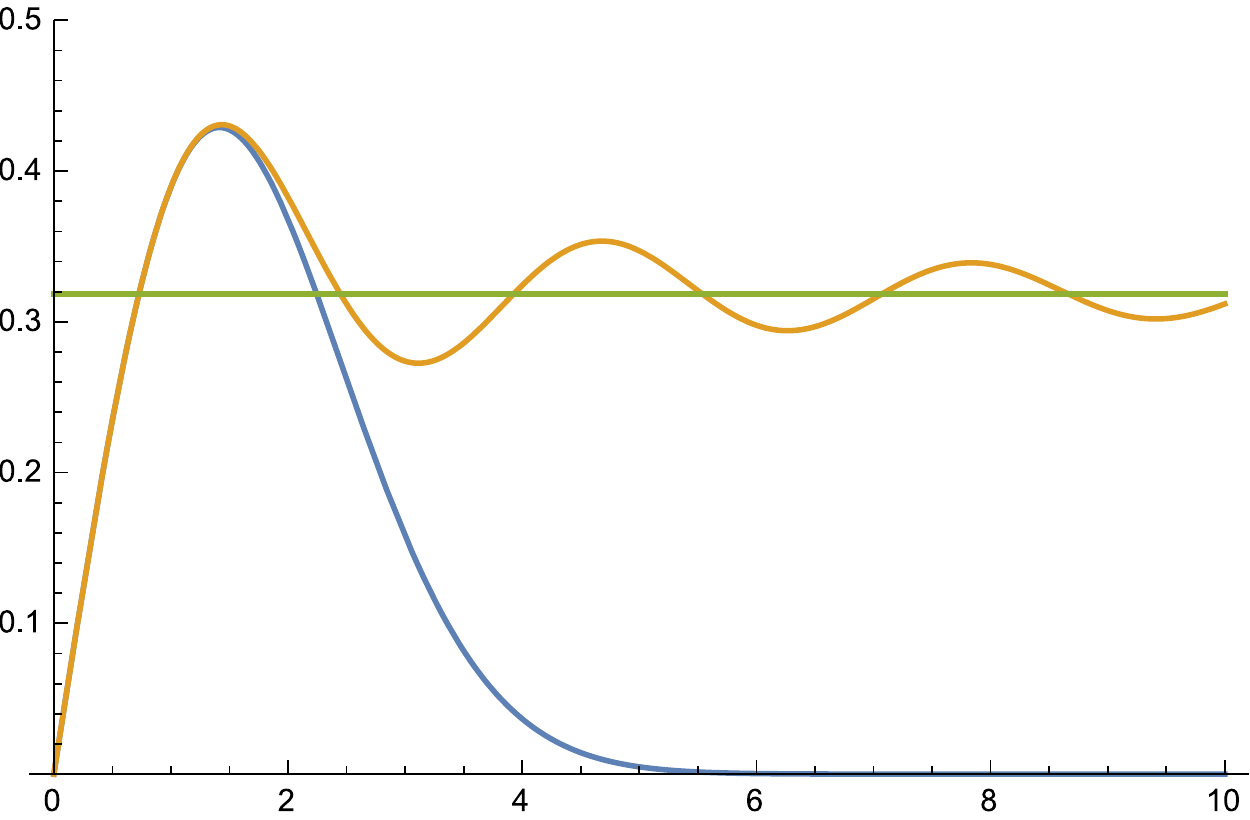}
\includegraphics[width=7cm]{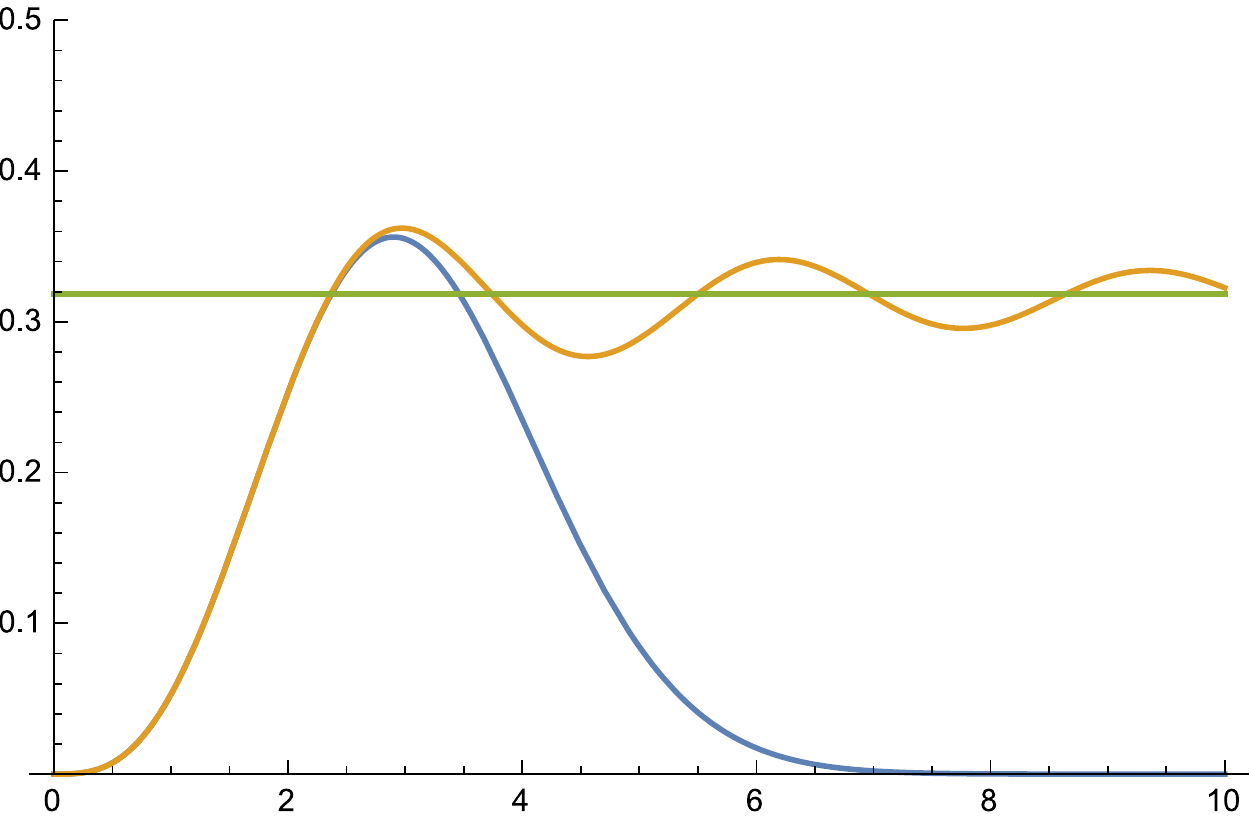}
\put(0.1,0.2){$\tilde{x}$}
\put(-7.1,0.2){$\tilde{x}$}
\put(-2.5,3.5){$\rho_s(\tilde{x})$}
\put(-10.8,3.5){$\rho_s(\tilde{x})$}
\put(-3.5,1.4){${\cal P}_1(\tilde{x})$}
\put(-11.5,1.4){${\cal P}_1(\tilde{x})$}
\end{picture}
\end{center}
\caption{The quenched microscopic spectral density from eq. (\ref{Besseldensity}) and the distribution of the smallest eigenvalue for $\nu=0$ (left) and $\nu=1$ (right) using eqs. (\ref{p1nu0Lim}) and (\ref{p1nu1Lim}). It can be nicely seen how the distribution of the smallest eigenvalues follows the microscopic density for small $\tilde{x}$. The local maxima of the density further to the right mark the locations of the second, third etc. eigenvalues, cf. \cite{Poul2000}. 
The limiting value  $\lim_{\tilde{x}\to\infty}\rho_s(\tilde{x})=1/\pi$ that equals the macroscopic density at the origin is marked as a horizontal line.
\label{fig:Besseldensity}}
\end{figure}

As the next step we will directly jump to the limiting kernel, the reason being two-fold. First, it is shorter to directly construct all correlation function form the kernel, eq. (\ref{Kernelmassive}), rather than constructing the kernel through the massive OP's first. Second and more importantly, the limit of the individual OP's does not necessarily exist. While here in the hard edge limit this is not the case, the well-known sine- and cosine-asymptotic of the Hermite polynomials in the bulk of the spectrum is only achieved after multiplying them with the square root of the weight function. 
We obtain for 
the limiting kernel from eq. (\ref{Kernelmassive}), after appropriately normalising and changing to Dirac eigenvalues, 
\begin{eqnarray}
{\cal K}_s^{[N_f]}(\tilde{y}_1,\tilde{y}_2)&\sim&
\lim_{\stackrel{\mbox{\small $N\to\infty$}}{\mbox{\small $x_1,x_2,m_f\to0$}}} (w(x_1)w(x_2))^{\frac12}K_N^{[N_f]}(x_1,x_2)\nonumber\\
&=&\frac{1}{(\tilde{y}_1^2-\tilde{y}_2^2)\prod_{f=1}^{N_f}\sqrt{(\tilde{y}_1^2+\tilde{m}_f^2)(\tilde{y}_2^2+\tilde{m}_f^2)}}
\frac{
\left|
\begin{array}{lll}
I_\nu(\tilde{m}_1)&\ldots&\tilde{m}_1^{N_f+1}I_{N_f+\nu+1}(\tilde{m}_1)\\
\vdots& & \vdots\\
I_\nu(\tilde{m}_{N_f})&\ldots&\tilde{m}_{N_f}^{N_f+1}I_{N_f+\nu+1}(\tilde{m}_{N_f})\\
J_\nu(\tilde{y}_1)&\ldots&(-\tilde{y}_1)^{N_f+1}J_{\nu+N_f+1}(\tilde{y}_1)\\
J_\nu(\tilde{y}_2)&\ldots&(-\tilde{y}_2)^{N_f+1}J_{\nu+N_f+1}(\tilde{y}_2)\\
\end{array}
\right|
}{\det_{1\leq f,g\leq N_f}[m_f^{g-1}I_{\nu+g-1}(\tilde{m}_f)]}.\ \ \ \ 
\label{KernelmassiveLim}
\end{eqnarray}
At $N_f=0$ it reduces to the standard Bessel-kernel:
\begin{equation}
{\cal K}_s(\tilde{x},\tilde{y})= -\frac{J_\nu(\tilde{x})\tilde{y}J_{\nu+1}(\tilde{y})-J_\nu(\tilde{y})\tilde{x}J_{\nu+1}(\tilde{x})}{\tilde{x}^2-\tilde{y}^2}= 
\frac{J_\nu(\tilde{x})\tilde{y}J_{\nu-1}(\tilde{y})-J_\nu(\tilde{y})\tilde{x}J_{\nu-1}(\tilde{x})}{\tilde{x}^2-\tilde{y}^2}\ .
\label{KBessel}
\end{equation}
Here we have given two equivalent forms of the same kernel using an identity \cite{GradshteynRyzhik} for $J$-Bessel functions, $2\nu J_\nu(x)=x(J_{\nu+1}(x)+J_{\nu-1}(x))$.
Following eq. (\ref{Rkb2}) we have already included the weight function into the limiting kernel above. While the exponential part cancels out due to $\exp[-x=-\tilde{y}^2/4N]\to1$ the remaining $\nu$-dependent part is contributing. We can thus immediately write down an example, the quenched microscopic density
\begin{equation}
\rho_s(\tilde{x})={\cal K}_s(\tilde{x},\tilde{x})=\frac{\tilde{x}}{2}\left( J_\nu(\tilde{x})^2-J_{\nu-1}(\tilde{x})J_{\nu+1}(\tilde{x})\right)\ .
\label{Besseldensity}
\end{equation}
We have omitted here the $\nu$ delta-functions of the exact zero-eigenvalues. They will enter the discussion later in Subsection \ref{WRMT} where they can be seen. The density is plotted in Figure \ref{fig:Besseldensity} for $\nu=0$ and 1.

The next step is to obtain the limiting distribution of the smallest eigenvalue. Because we change from Wishart to Dirac eigenvalues $s\to x^2$, we obtain the following scaling limit for the examples of the gap probability (\ref{E0nu0}) and (\ref{E0nu1}) at $N_f=0$,  using the same scaling for the Dirac eigenvalues as in eq. (\ref{RMTmass}):
\begin{eqnarray}
\nu=0:\ \ \lim_{\stackrel{\mbox{\small $N\to\infty$}}{\mbox{\small $x\to0$}}} E_0(s=x^2=\tilde{x}^2/(4N))&=& {\cal E}_0(\tilde{x})=e^{-\tilde{x}^2/4}\ ,
\label{E0nu0lim}\\
\nu=1:\ \ \lim_{\stackrel{\mbox{\small $N\to\infty$}}{\mbox{\small $x\to0$}}} E_0(s=x^2=\tilde{x}^2/(4N))&=& {\cal E}_0(\tilde{x})=e^{-\tilde{x}^2/4} I_0(\tilde{x})\ .
\label{E0nu1lim}
\end{eqnarray}
Note that the extra power of $N$ from the shift (\ref{gap-shift}) leads to the fact that here the exponential is not vanishing. The liming quenched gap probability 
for general $\nu$ follows from the completely degenerate massive partition function (\ref{ZRMTNfdeg}):
\begin{equation}
{\cal E}_0(\tilde{x})=e^{-\tilde{x}^2/4}\det_{1\leq i,j\leq \nu}[I_{i-j}(\tilde{x})]\ .
\label{Equenched}
\end{equation}
The limiting distribution of the corresponding smallest Dirac eigenvalue follows according to eq. (\ref{p1}), after changing to Dirac eigenvalues:
\begin{eqnarray}
\nu=0:\ \ {\cal P}_1(\tilde{x})&=&-\partial_{\tilde{x}}{\cal E}_0(\tilde{x})
= \frac{\tilde{x}}{2}e^{-\tilde{x}^2/4}\ ,
\label{p1nu0Lim}\\
\nu=1:\ \ {\cal P}_1(\tilde{x})&=&-\partial_{\tilde{x}} 
{\cal E}_0(\tilde{x})=
\frac{\tilde{x}}{2}e^{-\tilde{x}^2/4}I_2(\tilde{x})\ ,
\label{p1nu1Lim}
\end{eqnarray}
where for the last step we have used an identity for Bessel functions. The last two formulas are compared to the quenched Bessel density eq. (\ref{Besseldensity}) in Figure \ref{fig:Besseldensity}.
It turns out that for general $\nu$ the derivative of the determinant in eq. (\ref{Equenched}) can be written again as a determinant, following the ideas sketched after eq. (\ref{E0nu1}) that the 
distribution of the smallest eigenvalue can itself be expressed in term of an expectation value of characteristic polynomials, 
\begin{equation}
{\cal P}_1(\tilde{x})=\frac{\tilde{x}}{2}e^{-\tilde{x}^2/4}\det_{1\leq i,j\leq \nu}[I_{i-j+2}(\tilde{x})]\ ,
\label{p1quenched}
\end{equation}
see \cite{DN} for a detailed derivation, including masses.

It should be mentioned that alternative representations exist for the gap probability eq. (\ref{p1quenched}) that are based on Fredholm determinant analysis \cite{TWBessel}. They are given in terms of an exponential of an integral of the solution of a Painlev\'e V equation, which is valid for any real $\nu>-1$. For integer values of $\nu$ this expression reduces to the simpler forms that we have given here. Furthermore, in \cite{Peter94,Chen} representations in terms of hypergeometric functions with matrix argument were given that are valid for all $\beta>0$. In \cite{AVivo} the equivalence to the above expressions for $\beta=2$ was proved.

The distributions  that we have computed so far, as for example the density and smallest eigenvalue in Figure \ref{fig:Besseldensity} 
have been extensively compared to numerical solutions from lattice  QCD. 
For example the quenched densitiy in Figure \ref{fig:Besseldensity} was matched with QCD in \cite{Dam98a, Goc98}, whereas the quenched distribution of the smallest eigenvalue 
also given in Figure \ref{fig:Besseldensity} was compared in \cite{Edwards} to QCD for different values of $\nu$ (and also to other QFTs corresponding to $\beta=1,4$). 
A matching to unquenched results for the smallest eigenvalues can be found in \cite{Poul2000}, and we refer to \cite{TiloJac,Poul2002} for a more exhaustive discussion of the lattice results.

\section{Symmetries of QCD and its relation to RMT}\label{symm}

Apart from giving non-specialists some idea about the theory of Quantum Chromodynamics (QCD) we will have to answer three major questions in this section:

\begin{itemize}
\item How is the Dirac operator defined 
the spectrum of which can be described by RMT?

\item What are the global symmetries of QCD that determine the RMT to be used?

\item What is the limit that leads from QCD to this RMT?
\end{itemize}

Let us give first short answers that also indicate how this section will be organised. The eigenvalues of the QCD Dirac operator will be partly described by RMT, actually only its smallest eigenvalues, and we will introduce this operator first in Subsection \ref{QCD}. The anti-Hermiticity properties of the Dirac operator and the concept of chiral symmetry will allow us to introduce some formal aspects and the definition of the QCD partition function.  The breaking of chiral symmetry in Subsection \ref{XSB} then leads us to the first approximation of QCD, to chiral perturbation theory that only contains the lightest excitations, the Goldstone Bosons. In a last step in Subsection \ref{echPT} this interacting non-linear, effective QFT is simplified in the epsilon-regime, that agrees with RMT to leading order. 

In what follows it is important that we will talk about space-time coordinates $(x)=(\vec{x},t)=(x_1,x_2,x_3,x_4)$ with components $x_\mu$, $\mu=1,2,3,4$ equipped with the Euclidean metric $\delta_{\mu\nu}$ given by the Kronecker symbol\footnote{For this reason we don't need to distinguish between upper and lower indices as in Minkowski space-time.}.
This so-called Wick rotation from Minkowski to Euclidean metric is necessary for two reasons: only in this setting does the Dirac operator have defined anti-Hermiticity properties, and second only for QCD on a Euclidean space-time lattice numerical solutions of QCD can be easily performed to which we can compare our RMT. We use Einstein's summation conventions meaning  that Greek indices appearing twice are automatically summed over from 1 to 4.

\subsection{The Dirac operator and global symmetries of QCD}\label{QCD}

The QCD Dirac operator is a linear, matrix valued differential operator defined as follows:
\begin{equation}
D\!\!\!\!/=\gamma_\mu {D\!\!\!\!/}_{\mu} =\gamma_\mu\left(\partial_\mu+ig_s A_\mu(x)\right) = - {D\!\!\!\!/}^{\ \dag}\ ,
\label{Dirac}
\end{equation} 
where $\partial_\mu=\frac{\partial}{\partial x_\mu}$.
The $\gamma_\mu$ are the Dirac-gamma matrices in Euclidean space-time, satisfying the Clifford algebra
\begin{equation}
\{\gamma_\mu,\gamma_\nu\}=\gamma_\mu\gamma_\nu+\gamma_\nu\gamma_\mu=2\delta_{\mu\nu}. 
\label{Clifford}
\end{equation}
The curly brackets denote the anti-commutator. 
A standard representation as $4\times4$ matrices is given as follows in terms of the Pauli matrices $\sigma_k$, $k=1,2,3$:
\begin{equation}
\gamma_k=\left( 
\begin{array}{cc}
0& i\sigma_k\\
-i\sigma_k&0\\
\end{array}
\right)
,\ 
\gamma_4=\left( 
\begin{array}{cc}
0& 1_2\\
1_2&0\\
\end{array}
\right)
,\ 
\gamma_5=\gamma_1\gamma_2\gamma_3\gamma_4=
\left( 
\begin{array}{cc}
1_2&0\\
0&-1_2\\
\end{array}
\right).
\label{gammas}
\end{equation}
All matrices are Hermitian, $\gamma_\mu^\dag=\gamma_\mu$, including $\gamma_5$. This is only true in the Euclidean setting. Obviously one can construct projection operators as 
\begin{equation}
P_\pm=\frac12(1_4\pm\gamma_5), \ P_\pm^2=P_\pm,\ P_+P_-=P_-P_+=0,\ P_++P_-=1_4\ .
\label{projection}
\end{equation}
The real parameter $g_s$ in the Dirac operator is the coupling constant of the strong interactions, thus $g_s=0$ corresponds to the free Dirac operator without interactions. It acts on 4-vectors $\psi(x)$ called spinors in Minkowski space, due to their transformation properties there which we don't need to specify here. 
The real vector potential $A_\mu(x)\in SU_c(3)$ is an element of this Lie group for the three colours (the subscript $c$ stands for colour). It is a scalar with respect to the $\gamma_\mu$'s, just as the partial derivatives in eq. (\ref{Dirac}). The $\psi(x)$ come in 3 copies with colours blue $b$, green $g$ and red $r$, and so the interacting Dirac operator acts on the full vector $\Psi(x)=(\psi_b(x),\psi_g(x),\psi_r(x))$.   For each quark flavour up, down etc. we have such a vector $\Psi^q(x)$ with $q=1,2,\ldots N_f$. In the standard model of elementary particle physics we have $N_f=6$, but often we will keep $N_f$ fixed as a free parameter. 
We could have written the Dirac operator with tensor product notation to underline these structures, but the precise $SU_c(3)$ structure will not be crucial in the following.

We are now ready to state some important global symmetries of the Dirac operator. First, it follows from the algebra (\ref{Clifford}) that
\begin{equation}
0=\{D\!\!\!\!/\ ,\gamma_5\}\ \ \Rightarrow\ \ D\!\!\!\!/=
\left( 
\begin{array}{cc}
0&i{\cal W}\\
i{\cal W}^\dag&0\\
\end{array}
\right).
\label{Dchiral}
\end{equation} 
The Dirac operator is block off-diagonal also called chiral, where the symbol ${\cal W}$ still denotes a differential operator that depends on $x$ and $A_\mu$. Suppose we can find the eigenfunctions $\Phi_k(x)$ of the Dirac operator labelled by $k$:
\begin{equation}
D\!\!\!\!/\, \Phi_k(x)=i\lambda_k\Phi_k(x)\ ,
\label{eigenfunct}
\end{equation}
in a suitably regularised setting (e.g. a finite box).
Because of the anti-Hermiticity the eigenvalues $i\lambda_k\in i\mathbf{R}$ are purely imaginary. Furthermore, because of eq. (\ref{Dchiral}) multiplying equation (\ref{eigenfunct}) by $\gamma_5$ leads to different eigenfunctions $\gamma_5\Phi_k(x)$ with eigenvalues $-i\lambda_k$, provided that $\lambda_k\neq0$. Thus the non-zero eigenvalues  of the Dirac operator come in pairs $\pm i \lambda_k$.

The Euclidean QCD partition function or path integral as it is called in QFT can be formally written down as follows:
\begin{eqnarray}
{\cal Z}^{\rm QCD}&=&\int[dA_\mu]\int[d\Psi]\exp\left[ 
-\int d^4x \sum_{q=1}^{N_f}\overline{\Psi}^q(x)(D\!\!\!\!/+m_q)\Psi^q(x)-\int d^4x \frac12\mbox{Tr}(F_{\mu\nu}F_{\mu\nu})
\right]\nonumber\\
&=& \int[dA_\mu]\prod_{q=1}^{N_f}\det[D\!\!\!\!/+m_q]\exp\left[ -\int d^4x \frac12\mbox{Tr}(F_{\mu\nu}F_{\mu\nu})\right]\ .
\label{ZQCD}
\end{eqnarray}
Here $\overline{\Psi}^q(x)=\Psi^{q\,\dag}(x)\gamma_4$ is the Dirac conjugate and the $SU_c(3)$ field strength tensor is defined by the commutator $[{D\!\!\!\!/}_{\mu}, {D\!\!\!\!/}_{\nu} ]=-ig_sF_{\mu\nu}$.  The integration over $dA_\mu$ and $d\Psi$ are formal (path) integrals that we will not specify in the following, see standard textbooks on QCD for a discussion as \cite{Muta}. In the second equation however we have formally integrated out the quarks due to the following observation. For a complex $N$-vector $v$ and $N\times N$ matrix $B$ we have 
\begin{equation}
\int d^{2N}v\exp[-v^\dag B v]\sim \frac{1}{\det[B]}\ ,\ \ \int d^{2N}\psi\exp[-\psi^\dag B \psi]\sim {\det[B]}\ ,
\label{detintegrals}
\end{equation}
whereas in the second identity we have integrated over a complex vector $\psi$ of anticommuting variables. Integration and differentiation over such fermionic variables can be defined in a precise way, and we refer to chapter 7 in \cite{Handbook} for details. The reason for Fermions to be represented by anti-commuting variables is that in the canonical quantisation of free Dirac fields one has to use anti-commutators for these fields.

Comparing eq. (\ref{Dchiral}) and the second line of eq. (\ref{ZQCD}) with eq. (\ref{matrixrep}) we already get a first idea about the random matrix approximation of QCD, as it was suggested initially in \cite{SV93}: the Dirac operator from eq. (\ref{Dchiral}) is replaced by a constant matrix with the same block structure, ${\cal W}\to W$,
and the average over the fields strength tensor is replaced by a Gaussian average over $W$. We further learn in this approximation that the eigenvalues  $x_j=\lambda^2_k$ of the Gaussian random matrix $WW^\dag$ correspond to the squared eigenvalues of the Dirac operator rotated to the real line, and that the number of eigenvalues $N$ is proportional to the dimension of the QCD Dirac operator truncated in this way.
However, as we will explain in the next subsection the precise form of the limit from QCD to RMT was better understood later, representing a controlled approximation. 

Furthermore, in the case when all masses vanish $m_q=0$ the action in the exponent in the first line of eq. (\ref{ZQCD}) has an additional symmetry. Defining the vector $\Psi=(\Psi^1,\ldots\Psi^{N_f})$ as well as the projections $P_\pm\Psi=\Psi_{L/R}$ onto left (L) and right (R) chirality we can write the fermionic part of the action as
\begin{equation}
\sum_{q=1}^{N_f}\overline{\Psi}^q(x)D\!\!\!\!/\,\Psi^q(x)= \overline{\Psi}_R(x)D\!\!\!\!/\, \Psi_R(x)+\overline{\Psi}_L(x)D\!\!\!\!/\, \Psi_L(x) \ .
\label{Diracdecomp}
\end{equation}
It is invariant under the global rotations $\Psi_{L}(x)\to U_L\Psi_L(x)$ with $U_L\in U_L(N_f)$ and likewise for $R$. In contrast, even when all masses are equal, $m_q=m\ \forall q$, the mass term $m\overline{\Psi}\Psi=m(\overline{\Psi}_L\Psi_R+\overline{\Psi}_R\Psi_L)$ is only invariant under the diagonal transformation with $U_L=U_R$. In QCD the resulting global symmetry is 
$U_L(N_f)\times U_R(N_f)=U_V(1)\times U_A(1)\times SU_L(N_f)\times SU_R(N_f)$, with the  $U(1)$-factors called $U_V(1)$ for vector and $U_A(1)$ for axial-vector  split off. The latter is broken through quantum effects also called an anomaly, whereas the former remains unbroken through the mass term. This leads to the explicit chiral symmetry breaking pattern in QCD:
\begin{equation}
SU_L(N_f)\times SU_R(N_f)\to SU(N_f)\ .\label{XSBpattern}
\end{equation}

\subsection{Chiral symmetry breaking and chiral perturbation theory}\label{XSB}

In this subsection we will describe the first step of simplifying QCD to what is called its low-energy effective theory. We begin with the following observation. It is found that the QCD vacuum $|0\rangle$, the ground state with the lowest energy, breaks chiral symmetry {\it spontaneously}, in precisely the same way as the theory with equal masses in eq. (\ref{XSBpattern}) breaks it explicitly:
\begin{equation}
\Sigma=|\langle0|\overline{\Psi}\Psi |0\rangle| =|\langle0|\overline{\Psi}_L\Psi_R+\overline{\Psi}_R\Psi_L|0\rangle|\neq 0 \ .
\label{Sigma}
\end{equation}
The parameter $\Sigma$ is called the chiral condensate and it is the order parameter for the spontaneous breaking of the chiral symmetry (\ref{XSBpattern}). 
It turns out that for QCD the low energy phase with non-vanishing $\Sigma\neq0$ coincides with the phase\footnote{In toy models of QCD e.g. with two colours $SU_c(2)$ or different representations this is not necessarily the case.} where the quarks and the gluons, the carriers of the strong interaction represented by $A_\mu$,  are confined to colourless objects, see the phase diagram in Figure \ref{fig:phasediag}. In the phase with $\Sigma\neq0$ we need lattice QCD or other effective field theory description (such as RMT) because a perturbative expansion in the coupling constant $g_s$  will not work here, because the coupling is large.

An important consequence was drawn by Banks and Casher \cite{BC}, relating the global continuum density $\rho_{D\!\!\!\!/}(\lambda)$ of the Dirac operator eigenvalues in eq. (\ref{eigenfunct}) at the origin, suitably regularised in a finite volume $V$, to $\Sigma$:
\begin{equation}
\rho_{D\!\!\!\!/}(\lambda\approx0)=\frac{1}{\pi}\Sigma V+ |\lambda|\frac{\Sigma^2}{32\pi^2 F_\pi^4 N_f}(N_f^2-4) + o(\lambda)\ .
\label{BCSS}
\end{equation} 
Here we added the second order term to eq. (\ref{BCrel}) derived by Stern and Smilga \cite{SS}, containing the pion decay constant $F_\pi$ as an additional parameter. The first term leads to the following interpretation: at the origin the global density is constant, and thus the average distance between eigenvalues there is $\lambda_k\sim 1/V$. This is very different from free particles in a box where $\lambda\sim 1/L$ with $V=L^4$ in 4 Euclidean dimensions. One can thus say that the smallest Dirac operator eigenvalues make an important contribution to build up the chiral condensate $\Sigma\neq0$, by piling up very closely spaced at the origin.
We are thus lead to define rescaled, dimensionless eigenvalues and masses, as they appear on the same footing inside the determinant  in eq. (\ref{ZQCD}), together with a rescaled microscopic density of the Dirac operator $\rho_s(\hat{\lambda})$:
\begin{equation}
\hat{\lambda}_k= \Sigma V \lambda_k,\ \hat{m}_f= \Sigma V m_f, \ \rho_s(\hat{\lambda})=\lim_{V\to\infty}\frac{1}{\Sigma V}\rho_{D\!\!\!\!/}(\hat{\lambda}/\Sigma V)\ .
\label{microlim}
\end{equation} 
This is called the microscopic limit of QCD. It has to be compared e.g. to eq. (\ref{Besseldensity}) for $N_f=0$, after identifying rescaled RMT eigenvalues $\tilde{y}_j=\hat{\lambda}_j$ and masses $\tilde{m}_f=\hat{m}_f$ from eq. (\ref{RMTmass}).

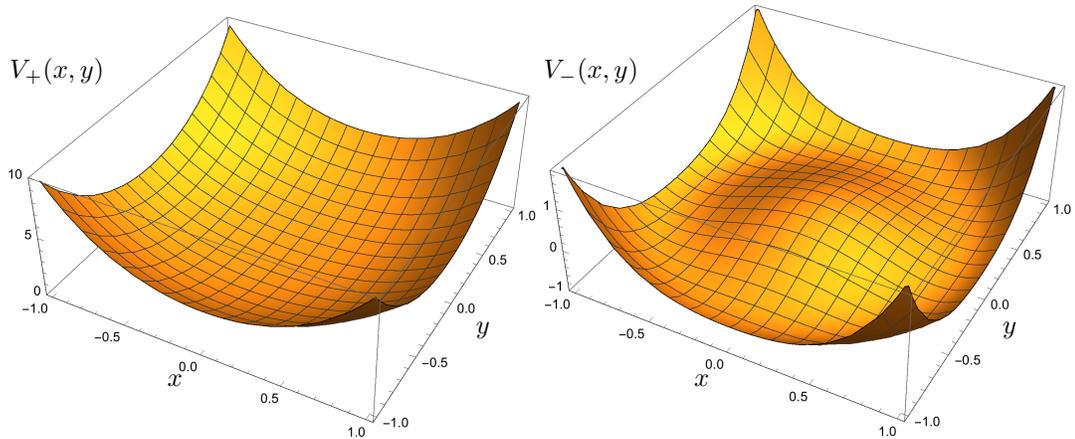
\begin{figure}[h]
\unitlength1cm
\begin{center}
\begin{picture}(12.2,5.5)
\includegraphics[width=7cm]{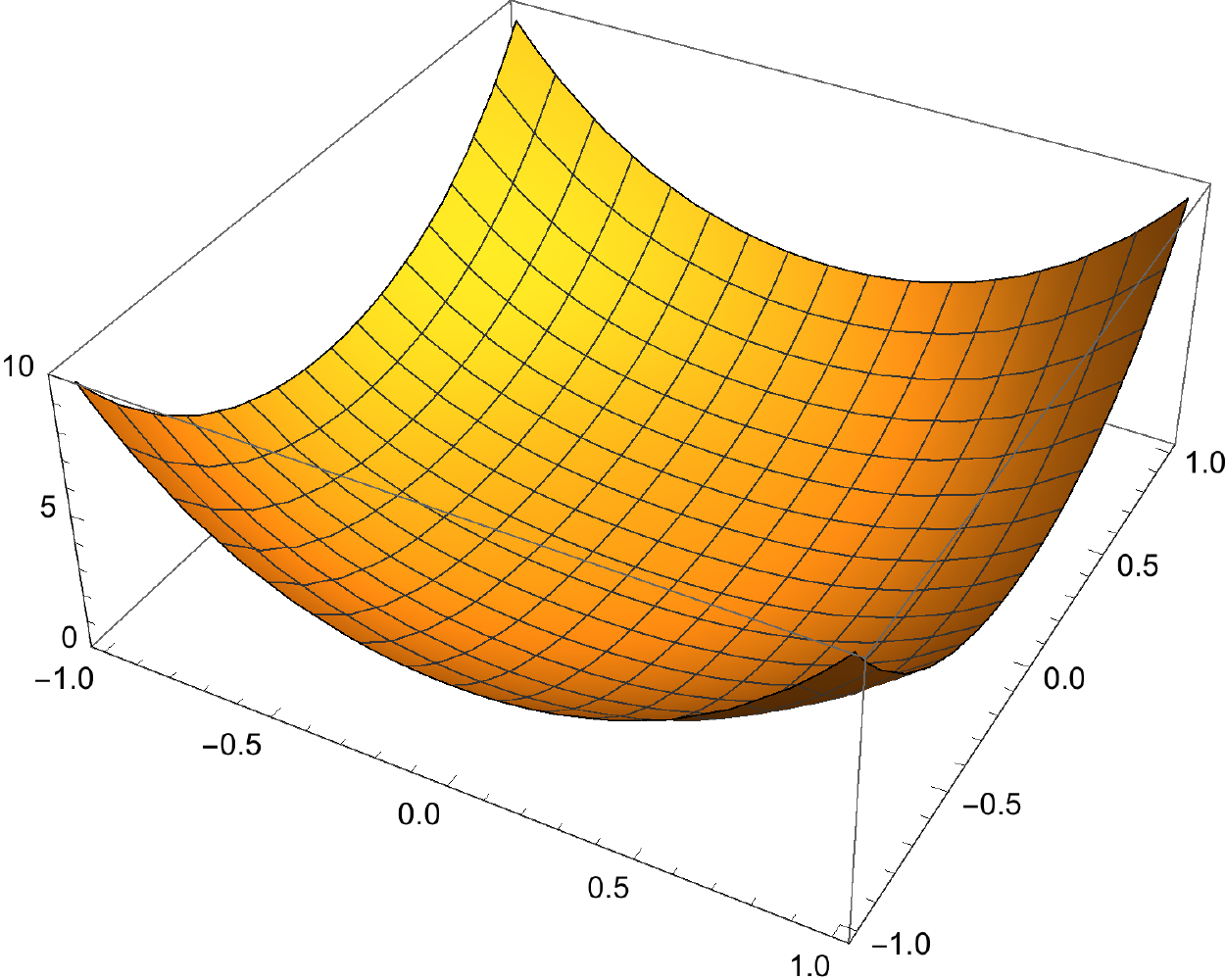}
\includegraphics[width=7cm]{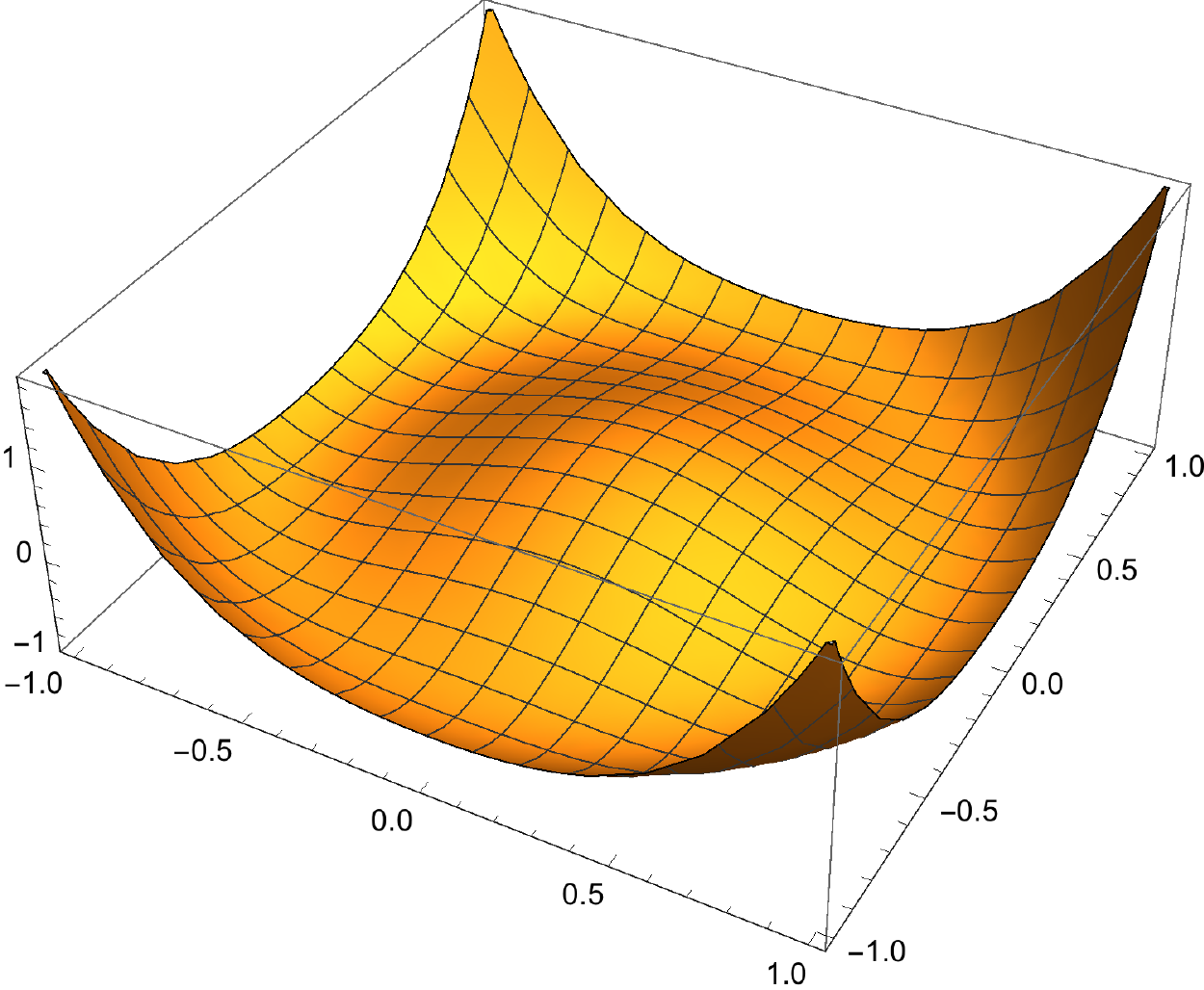}
\put(-5.0,0.7){$x$}
\put(-12.0,0.7){$x$}
\put(-7.0,4.8){$V_-(x,y)$}
\put(-0.9,1.4){$y$}
\put(-7.9,1.4){$y$}
\put(-14.1,4.8){$V_+(x,y)$}
\end{picture}
\end{center}
\caption{The potentials $V_\pm(x,y)=\pm 2 (x^2+y^2)+ 1.5 (x^2+y^2)^2$ before symmetry breaking ($V_+$, left) and after breaking ($V_-$, right), obtained simply by switching the sign of the quadratic term.
\label{fig:Goldstone}}
\end{figure}

The breaking of a global continuous symmetry in any theory leads to Goldstone Bosons. This can be visualised by analogy in Figure \ref{fig:Goldstone}, showing the classical potential of a particle before (left) and after symmetry breaking (right). In the left picture with a convex potential any excitation to make a particle move in the potential costs energy. After symmetry breaking the curvature at the origin has changed. Consequently there now exists the possibility to excite a mode in angular direction along the valley of the potential, without costing potential energy. Modes that do need energy to be excited still exist in radial direction.

In our analogy particles that require energy to be excited correspond to massive particles. Consequently after symmetry breaking there appear massless modes - the Goldstone Bosons - and modes that remain massive. If in QCD chiral symmetry was an exact symmetry we would expect to observe exactly massless Goldstone Bosons in the phase with broken symmetry. However, it is only an approximate symmetry due to $m_q\neq0$, leading to approximately massless Bosons after symmetry breaking. These are the 3 light pions observed in nature, $\pi^\pm$ and $\pi^0$, with masses of about 135 Mev compared to the heavier non-Goldstone particles with masses of 800 -1000 Mev. Recall that in the confined phase we only observe colourless bound states made from two quarks, Mesons, which are Bosons, from three quarks, Baryons, which are Fermions,  or more\footnote{During the delivery of these lectures in July 2015 the discovery of the Pentaquark was announced at CERN.}. The pions are unstable and decay mostly into a muon and its anti-neutrino for $\pi^+$, the corresponding anti-particles for $\pi^-$, or 2 photons for $\pi^0$.

 The number of Goldstone Bosons equals the dimension of the coset  $SU_L(N_f)\times SU_R(N_f) / SU(N_f)$ of the groups before and after breaking, which is $N_f^2-1$, the dimension of $SU(N_f)$ in our case. To consider the pions as Goldstone Bosons thus means that we only keep the lightest quarks up and down, so that $N_f=2$ in our theory. For many purposes at low energy this is a good approximation. Sometimes also the strange quark is included, with $N_f=3$ leading to consider the 8 lightest particles as Goldstone Bosons.
 
 If we neglect all non-Goldstone Bosons in our theory - as they will not be relevant at low enough energies - and formally integrate them out we arrive at the low-energy effective theory of the Goldstone Bosons,
 \begin{equation}
 {\cal Z}_{\rm chPT}=\int[dU]\exp\left[-\int d^4x \frac{F_\pi^2}{4}\mbox{Tr}\left({\partial_\mu U(x)}{\partial_\mu U(x)^\dag}\right)
+\int d^4x\frac{\Sigma}{2}\mbox{Tr}\left(M(U(x)+U^\dag(x))\right)
 \right]
\label{ZchPT} 
\end{equation}  
where
\begin{equation}
U(x)=U_0\exp\left[i\frac{\sqrt{2}}{F_\pi}
\pi^b(x)t^b\right], \  M=\mbox{diag}(m_1,\ldots,m_{N_f}).
\label{U(x)}
\end{equation} 
Here the implicit sum over $b$ runs from 1 to $N_f^2-1$.
For $N_f=2$ the 3 fields $\pi^b(x)$ actually denote the pion fields, with $t^3=\sigma_3$ and $t^{\pm}=\sigma^\pm=\sigma^1\pm\sigma^2$. 
In eq. (\ref{U(x)}) we have already split off the constant, $x$-independent modes  $U_0$ of the fields $\pi^b(x)$ for later convenience. Usually at this stage $U_0$ is set to the identity.
Eq. (\ref{ZchPT}) represents the partition function or path integral of chiral perturbation theory (chPT), which replaces that of QCD eq. (\ref{ZQCD}) in the low-energy regime.
In eq. (\ref{ZchPT}) we only give the leading order (LO) Lagrangian, all higher order terms in powers of $U(x)$ and its derivatives that are invariant exist and we will come back to the question when they contribute in the next section. For more details about  chPT we refer to \cite{Scherer}.

ChPT is an interacting, non-linear QFT.
When expanding the exponential in eq. (\ref{U(x)}) in powers of $\pi^b(x)$ and considering only quadratic terms we obtain the standard Lagrangian for scalar fields that leads to the Klein-Gordon equation. From that expansion we can make the following identification for the masses $M_\pi$ of the pion fields $\pi^b(x)$, the Gell-Mann--Oakes--Renner (GOR) relation:
\begin{equation}
F_\pi^2M_\pi^2=\Sigma (m_1+\ldots+m_{N_f})\ .
\label{GOR}
\end{equation}
The QFT in eq. (\ref{ZchPT}) is not yet related to RMT, and we turn to this limit in the next subsection.

\subsection{The limit to RMT: the epsilon-regime}\label{echPT}

The so-called epsilon-regime of chPT (echPT) was introduced by Gasser and Leutwyler \cite{GL87}  before a RMT of QCD was constructed. Their motivation was to introduce a regime where analytic computations in chPT were feasible, in particular to understand the mass dependence of the partition function and the role of topology. 
In particular Leutwyler and Smilga later computed the partition function analytically in this regime and deduced so-called sum rules for the Dirac operator eigenvalues \cite{LS}.  
Working in a finite volume $V=L^4$ of linear size $L$, the authors of  \cite{GL87} introduced the following scaling:
\begin{equation}
\epsilon= L^{-1}\ \Rightarrow\ V\sim \epsilon^{-4},
\ \partial_\mu\sim  \epsilon,\ \pi^b(x)\sim\epsilon\ .
\label{epsilon}
\end{equation}
This scaling implies in particular
\begin{equation}
\Sigma V m_f={\cal O}(1),\ \int d^4x \partial_\mu\pi^b(x)\partial_\mu\pi^b(x) = {\cal O}(1)
\label{exponenteps}
\end{equation}
where the former follows from the microscopic scaling limit of the Banks-Casher relation (\ref{microlim}), and the latter ensures that the kinetic term of the LO action is dimensionless. The physical interpretation of the epsilon-regime follows from the scaling of the pion mass, which is implied from multiplying the GOR relation eq. (\ref{GOR}) with $V$. It has to hold that $M_\pi\sim L^{-2}$, such that both sides are of the order ${\cal O}(1)$. In other words, the Compton wavelength of the pion fields is much larger than the size of the box:
\begin{equation}
M_\pi^{-1}\sim L^2\gg L\ , 
\label{Compton}
\end{equation}
which is clearly an unphysical regime. Normally one would consider this to be undesirable, because in order to observe physical, propagating pions one would choose a scaling such that the pions do fit into the box $V$, with $M_\pi^{-1}< L$. In the so-called $p$-regime both momenta and pion mass are scaled in the same way, $\partial_\mu,\ M_\pi\sim  \epsilon$, where such an inequality can be satisfied. We will come back to this issue very briefly below.
In the epsilon-regime introduced above we obtain the following  limit \cite{GL87}
\begin{equation}
\lim_{\mbox{$\epsilon$-regime}}{\cal Z}_{\rm chPT}= \int [d\pi] \exp\left[-\int d^4x \frac12\partial_\mu\pi^b(x)\partial_\mu\pi^b(x) \right]\int_{SU(N_f)}dU_0\exp\left[\frac12 \Sigma V\mbox{Tr}(M(U_0+U_0^\dag))+{\cal O}(\epsilon^2)\right].
\label{ZechPT}
\end{equation}
Here the propagating modes $\pi^b(x)$ completely factorise from the zero-modes $U_0$ and contribute to the partition function as a free Gaussian field theory - which is just a constant factor. The reason that we do not expand in the constant modes of the pions as well, but keep them as a full matrix $U_0$ as in eq. (\ref{U(x)}) is that their quantum fluctuations are of ${\cal O}(1)=V/M_\pi^2$, see \cite{Poul2011} for a more detailed discussion. All higher order terms including those not given in the chiral Lagrangian (\ref{ZchPT}) are subleading in this limit. 
It is in this limit that the mass-dependence of the limiting partition function eq. (\ref{ZechPT}) and of RMT agree, and in fact also all Dirac operator eigenvalue correlation functions. Before we make this statement more precise let us add further comments. First, it is not only possible to make analytical computations in the epsilon-regime, but also to compare them to lattice  QCD. On the lattice the simulation parameters can be freely chosen, such that the epsilon-regime is reached and a quantitative comparison is possible. 
Second, it is possible to compute so-called one or higher loop corrections to the partition function (and correlation functions) eq. (\ref{ZechPT}). For example writing down the  ${\cal O}(\epsilon^2)$ term explicitly, expanding the exponential  and using Wick's theorem with the finite volume propagator corresponds to computing the one-loop corrections. Third, one may perform also perturbative computations in the physical $p$-regime, or even chose an extrapolation between the two regimes, cf. \cite{Poul-Hide,Poul2011} for a discussion and references.

Let us now make the map between RMT and echPT more precise. For that purpose it is convenient to introduce an extra term in the QCD Lagrangian, the so-called theta-vacuum term containing the dual field strength tensor $\tilde{F}_{\rho\sigma}\sim\varepsilon_{\rho\sigma\mu\nu}F^{\mu\nu}$. It is defined using the totally antisymmetric epsilon-tensor $\varepsilon_{\rho\sigma\mu\nu}$ in 4 dimensions. The extra term reads in appropriate normalisation:
\begin{equation}
-i\theta\int d^4x \mbox{Tr}\tilde{F}_{\rho\sigma}{F}_{\rho\sigma}=-i\nu\theta\ .
\label{thetaterm}
\end{equation}
It is a topological term, and here $\nu$ is the number of left handed $(L)$ minus right handed $(R)$ zero modes of the Dirac operator eq. (\ref{Dirac}) - which so far has not made an appearance in this section about QCD. Due to the Atiah-Singer index theorem $\nu$ is equal to the winding number of the gauge fields $A_\mu$ which is also a topological quantity. In the partition function eq. (\ref{ZQCD}) we have thus implicitly summed over all the topologies, which can now be written as follows:
\begin{equation}
{\cal Z}^{\rm QCD}(\theta)=\sum_{\nu=-\infty}^\infty e^{-i\nu\theta}{\cal Z}^{\rm QCD}_\nu\ .
\label{fixtopo}
\end{equation}
Here ${\cal Z}^{\rm QCD}_\nu$ denotes the QCD partition function at fixed topology. Inverting this equation it can be written as 
\begin{equation}
{\cal Z}^{\rm QCD}_\nu=\frac{1}{2\pi}\int_{-\pi}^\pi d\theta  e^{+i\nu\theta}{\cal Z}^{\rm QCD}(\theta)\ .
\end{equation}
What is the advantage? In RMT in the previous section we always had a fixed $\nu\in\mathbf{N}$ - to which we should now compare for non-negative values\footnote{The QCD partition function is thought to be symmetric with respect to $\nu$.}. Furthermore, the unitary group integral over $SU(N_f)$ that we found in the epsilon-regime eq. (\ref{ZechPT}) can now be promoted to an integral over the full unitary group, according to
\begin{equation}
\int_{-\pi}^\pi\frac{d\theta}{2\pi} e^{i\nu\theta}\int_{SU(N_f)}dU_0 = \int_{U(N_f)}dU_0\det[U_0]^\nu
\label{groupint}
\end{equation}
Defining the rescaled quark masses in the epsilon-regime as follows,
\begin{equation}
\hat{M}=\mbox{diag}(\hat{m}_1,\ldots,\hat{m}_{N_f})=\Sigma V M\ ,
\label{hatM}
\end{equation}
we obtain the limiting relation between the mass-dependent part of the QCD partition function in the epsilon-regime at fixed topology and RMT:
\begin{eqnarray}
{\cal Z}_{\rm echPT}^{[N_f]}(\{\hat{m}\})&=&\int_{U(N_f)} dU_0\det[U_0]^\nu\exp\left[\frac12 \mbox{Tr}(\hat{M}(U_0+U_0^\dag))\right]\nonumber\\
&=&
2^{\frac{N_f(N_f-1)}{2}}\prod_{j=0}^{N_f-1}\Gamma(j+1)
\frac{\det_{1\leq i,j\leq N_f}[\hat{m}_i^{j-1}I_\nu(\hat{m}_i)]}{\Delta_{N_f}(\{\hat{m}^2\})}=
{\cal Z}^{[N_f]}(\{\hat{m}\})\ .
\label{LSint}
\end{eqnarray}
In the last equality with the limiting RMT partition function from eq. (\ref{ZRMTNf}) we have identified rescaled random matrix and QCD quark masses $\tilde{m}_f=\hat{m}_f$ for all flavours.
The computation of the integral over the appropriately normalised Haar measure of the unitary group $U(N_f)$ in eq. (\ref{LSint}) goes back to \cite{Brower}, see also \cite{JSV} and \cite{Baha} for our normalisation. A concise derivation in terms of group characters was presented in \cite{Baha}. In the simplest case with $N_f=1$ it is easy to see that:
\begin{equation}
{\cal Z}_{\rm echPT}^{[N_f=1]}(\hat{m})=\int_{-\pi}^\pi\frac{d\theta}{2\pi} e^{i\nu\theta}\exp\left[ \frac12 \hat{m}(e^{i\theta}+e^{-i\theta})\right]=I_\nu(\hat{m})\ ,
\label{ZechPTNf1}
\end{equation}
which is one of the standard definitions of the modified Bessel-function. Furthermore, we can also verify the completely degenerate case $m_f=m,\ \forall\ f$, by  applying the Andr\'eief formula (\ref{Andreiev}). Diagonalising $U_0$ by a unitary transformation $U\in U(N) /U(1)^N$, $U_0=U\mbox{diag}(e^{i\theta_1},\ldots,e^{i\theta_{N_f}})U^\dag$, we obtain again the squared Vandermonde determinant as a Jacobian:
\begin{eqnarray}
&&\int_{U(N_f)} dU_0\det[U_0]^\nu\exp\left[\frac{\hat{m}}{2}\mbox{Tr}(U_0+U_0^\dag)\right]=const. \prod_{j=1}^{N_f}\left(\int_{-\pi}^\pi\frac{d\theta_j}{2\pi}\exp[i\nu\theta_j+\hat{m}\cos(\theta_j)]\right) |\Delta_{N_f}(\{e^{i\theta}\})|^2\nonumber\\
&=&N_f!\ const.\det_{1\leq j,k\leq N_f}\left[ \int_{-\pi}^\pi\frac{d\theta}{2\pi}\exp[i(\nu+k-j)\theta+\hat{m}\cos(\theta)]\right]=\det_{1\leq j,k\leq N_f}[I_{\nu+k-j}(\hat{m})]\ .
\label{Zdeg}
\end{eqnarray}
It agrees with eq. (\ref{ZRMTNfdeg}) from RMT, after appropriately normalising. We observe that both for degenerate and non-degenerate masses the partition functions with $N_f$ flavours eqs. (\ref{LSint}) and (\ref{Zdeg}) are determinants of single flavour partition functions with $N_f=1$.

The matching of the limiting RMT partition function and the group integral from the epsilon-regime of chPT at fixed topology is only part of their equivalence. In particular this matching
does not imply that all Dirac operator eigenvalue correlation functions agree. For example at $N_f=1$ the echPT partition function agrees for all three symmetry classes $\beta=1,2,4$ 
\cite{JacSmilga}, while their densities differ.
At first sight it is not clear how to access the Dirac operator spectrum, once we have moved from the quarks to the Goldstone Bosons as fundamental low energy degrees of freedom. However, looking at the alternative construction of the spectral density via the resolvent eq. (\ref{resolvent}) helps. Consider adding an auxiliary pair of quarks with masses $x$ and $y$ respectively to the QCD Lagrangian eq. (\ref{ZQCD}). If we arrange for the second quark to be bosonic, we obtain the following average with respect to the unperturbed QCD partition function, at fixed topology:
\begin{equation}
\lim_{\mbox{$\epsilon$-regime}}
\left\langle \frac{\det[x-D\!\!\!\!/\ ]}{\det[y-D\!\!\!\!/\ ]}\right\rangle_\nu^{\rm QCD}= \frac{\int_{U(N_f+1|1)} dU_0\ \mbox{sdet}[U_0]^\nu\exp\left[\frac12 \mbox{sTr}(\hat{{\cal M}}(U_0+U_0^\dag))\right]}{ {\cal Z}_{\rm echPT}^{[N_f]}(\{\hat{m}\})}\ ,
\label{QCDresolvent}
\end{equation}
with $\hat{{\cal M}}=\mbox{diag}(\hat{m}_1,\ldots,\hat{m}_{N_f},\hat{x};\hat{y})$,
in the limit of the epsilon-regime. Because of the bosonic nature of the second auxiliary quark in the denominator the integral is now over the supergroup $U(N_f+1|1)$ with the corresponding supersymmetric trace and determinant, sTr and sdet respectively, see e.g. chapter 7 of \cite{Handbook}. Its evaluation and differentiation to obtain the resolvent has to be done carefully. In \cite{DOTV} to where we refer for details it was shown in this fashion that the microscopic density eq. (\ref{Besseldensity}) and its extension to arbitrary $N_f$ follows from eq. (\ref{QCDresolvent}), thus establishing the equivalence to RMT on the level of the microscopic density.  In order to know the distribution of the smallest eigenvalues the knowledge of all density correlation functions is necessary, see eq. (\ref{expand}), as was pointed out in \cite{AD03}. The equivalence proof for all $k$-point density correlation functions was achieved in \cite{BA} by directly matching the $k$-point resolvent generating functions with $2k$ additional ratios of quark determinants of both theories, using superbosonisation techniques. 

A few comments regarding universality and corrections to the RMT regime are in place here. Already the fact that two apparently different theories as echPT and RMT yield the same Dirac operator correlation functions is a striking result which is based on symmetries and universality. It is thus surprising that even when taking into account one-loop corrections to echPT the same universal 
correlation functions were found in \cite{CT1}. The only corrections to be made are subsumed in a modification $\Sigma\to\Sigma_{\rm eff}=\Sigma(1+C_1/\sqrt{V})$, and $F_\pi\to F_{\rm eff}=F_\pi(1+C_2/\sqrt{V})$, where $C_1$ and $C_2$ are constants that depend on $N_f$, $F_\pi$ and the geometry of the discretisation, see e.g. \cite{ABL,DDF} for details and references. So far $F_\pi$ did not appear in eigenvalue correlation functions, but later when including a chemical potential it will. This modification
 means that the chiral condensate (and pion decay constant) and thus the rescaled masses in eq. (\ref{hatM}) and eigenvalues simply get renormalised accordingly. Only two-loop corrections within  echPT contain non-universal terms that lead out of the RMT universality class \cite{CT2}. 
 
However, even at the LO the agreement between the RMT and echPT rapidly breaks down when going to higher energies, meaning to higher rescaled Dirac operator eigenvalues. 
At some point the fluctuating modes will start to contribute, and eventually also the chPT approach breaks down as contributions from full QCD will appear. The scale where this happens has been called Thouless energy \cite{JacJames,Janik} in analogy to applications in condensed matter physics, to where we refer for details. The point where this happens scales with $1/\sqrt{V}$ and $F_\pi^2$. Beyond this scale one sees a rise of the global spectral density as in eq. (\ref{BCSS}), away from the RMT prediction. 
The scaling of the Thouless energy has been confirmed from lattice data in a different theory, in QFT with two colours \cite{BB}. For the same theory deep in the bulk regime the local statistics was seen to agree again with RMT bulk statistics of the GSE \cite{GuhrMa}, which is the non-chiral $\beta=4$ symmetry class.

Here we have already mentioned one of  the two different symmetry classes corresponding to $\beta=1$ and 4 in eq. (\ref{partitionfunct}). 
They are also relevant but for different QFTs other than QCD, e.g. with only two colours $SU_{c}(2)$. These have chiral symmetry patters different from  eq. (\ref{XSBpattern}) and we refer to \cite{TiloJac} for a detailed discussion. The matching with the 3 chiral RMTs in eq. (\ref{partitionfunct}) was initially proposed in \cite{Jac3fold}. While the agreement with lattice simulations of the respective theories (cf. \cite{Poul2002}) leave little doubt that these RMT are equivalent to the epsilon-regime of the corresponding chPT, such an equivalence has not been established even on the level of the partition function for general $N_f$. This is due to the more complicated structure of the corresponding group integral, see however \cite{JacSmilga} and \cite{JacDominique} for a discussion.

Let us briefly comment also on the influence of the dimension. In 3 space-time dimensions there is no chiral symmetry, and the RMT for QCD is given by the GUE instead \cite{JacQCD3}, including mass terms. In 2 dimensions which again has chiral symmetry the classification is much richer, and we refer to \cite{JacMario2D} for details. This makes contact with the symmetry classification of topological insulators \cite{Ludwig} and the so-called Bott-periodicity in any dimension.

\section{Recent developments}\label{recent}

In this section we will cover some recent developments in the application of RMT to QCD, where the detailed dependence of the partition function and Dirac operator eigenvalue 
correlation functions on finite lattice spacing $a$ or chemical potential $\mu$ are computed. In the first Subsection \ref{WRMT} we will study the influence of a finite space-time lattice on the spectrum as it is expected to be seen in numerical simulation of QCD on a finite lattice far enough from the continuum. The corresponding RMT and effective field theory that we will consider go under the name of Wilson (W)RMT as introduced in \cite{DSV}, and Wilson (W)chPT, see \cite{Sharpe} for a standard reference. We will only sketch the ideas beginning with the symmetries and WchPT, and then write down the joint density of eigenvalues of the Hermitian Wilson Dirac operator. Its solution and its non-Hermitian part will be referred to the literature.

The addition of a chemical potential term for the quarks to the Dirac operator renders its eigenvalues complex. The solution  of the corresponding RMT using OP in the complex plane is sketched in the second Subsection \ref{RMTmu} where we will give some details. The first application of such a RMT was already introduced by Stephanov in 1996 \cite{Mischa} to explain the difference between the quenched and unquenched theory. The development of the corresponding theory of OP in the complex plane started later in \cite{A01} and \cite{James}. It is already partly covered in the Les Houches lecture notes from 2005 \cite{Jac2005}, for a slightly more recent review on RMT of QCD with chemical potential see \cite{A07}. Here we will comment on  some recent developments relating to products of random matrices, see \cite{AIp} for a detailed review (and the lectures of Y. Tourigny in this volume).

\subsection{RMT and QCD at finite lattice spacing - the Wilson Dirac operator}\label{WRMT}

The discretisation of derivatives on a space-time lattice is not a unique procedure, which also applies to the Dirac operator in eq. (\ref{Dirac}). We would thus first like to motivate why in this subsection we will study the modification proposed by Wilson, the Wilson Dirac operator
\begin{equation}
D_W=D\!\!\!\!/ +a\Delta \neq -D_W^\dag\ ,
\label{WDirac}
\end{equation} 
which is no longer anti-Hermitian.
Here $a$ is the lattice spacing and $\Delta$ denotes the Laplace operator which is Hermitian. Of course both operators still have to be discretised. 
We cannot possibly give justice to the vast literature on this subject  and will only focus on aspects relevant for the application of RMT.
The reason why Wilson proposed to add the Laplacian is the so-called doubler problem. In the continuum with $a=0$ the relativistic energy-momentum relation for a particle with mass $M$, energy $E$, and 4-momentum $k_\mu$ reads
\begin{equation}
E^2-M^2=\sum_{\mu=1}^4 k^2_\mu\ .
\label{Emom}
\end{equation}
Here we have spelled out the sum explicitly. Particles satisfying this relation are called on-shell (on the energy shell). After standard discretisation this relation becomes 
\begin{equation}
E^2-M^2=\sum_{\mu=1}^4 \frac{\sin(k_\mu a)^2}{a^2}\ ,
\label{Emoma}
\end{equation}
which in the limit $a\to0$ leads back to eq. (\ref{Emom}). In contrast to the continuum relation, here with any $k_\mu$ satisfying this equation also $(k_1-\frac{\pi}{a},k_2,k_3,k_4)$, 
$(k_1,k_2-\frac{\pi}{a},k_3,k_4)$ etc. fulfil eq. (\ref{Emoma}). In total we have $2^4=16$ possibilities to add $-\pi/a$ to the components of $k_\mu$, which all correspond to the same on-shell particle. This is the doubler problem.
Wilson's modification to add the Laplacian to the Dirac operator  amounts to adding the term $\gamma_\mu \sin(k_\mu a)$ to the right hand side of eq. (\ref{Emoma}). In the limit $a\to0$ this makes the extra 15 particles heavy and thus removes the doublers. The modification comes with a prices, as the Laplacian explicitly breaks chiral symmetry. We will not discuss other possibilities here, for example in \cite{James-staggered} for a RMT of the so-called staggered  Dirac operator, but rather stick to Wilson's choice.
Despite the non-Hermiticity of $D_W$ eq. (\ref{WDirac}) it satisfies the so-called $\gamma_5$-Hermiticity:
\begin{equation}
D_W^\dag=-D\!\!\!\!/ +a\Delta=\gamma_5 D_W\gamma_5\ .
\label{ga5}
\end{equation}
Here we have simply used that $(\gamma_5)^2=1_4$ and the anti-commutator from eq. (\ref{Dchiral}). Consequently one can define the following Hermitian operator called $D_5$
\begin{equation}
D_5=\gamma_5(D_W+m)=D_5^\dag\ ,
\label{D5}
\end{equation}
where traditionally the quark mass $m$ is added to the definition. Looking at eq. (\ref{Dchiral}) for the Dirac operator and eq. (\ref{gammas}) for $\gamma_5$ we can immediately give the chiral block structure of $D_W$ and $D_5$:
\begin{equation}
D_W=
\left( 
\begin{array}{cc}
a{\cal A}&i{\cal W}\\
i{\cal W}^\dag&a{\cal B}\\
\end{array}
\right)\ ,\ \ 
D_5=
\left( 
\begin{array}{cc}
m&i{\cal W}\\
-i{\cal W}^\dag&-m\\
\end{array}
\right)+
a
\left( 
\begin{array}{cc}
{\cal A}&0\\
0&-{\cal B}\\
\end{array}
\right).
\label{DWchiral}
\end{equation}
Here ${\cal A}$ and ${\cal B}$ are Hermitian operators. 
Note the change in sign in the mass term in the lower right block of $D_5$. As previously for $a=0$ this immediately leads to a good Ansatz to make for a WRMT, as it was made in \cite{DSV}. Before we turn to analyse this in some detail let us directly go to the epsilon-regime of QCD in the Wilson formulation at fixed topology, WechPT. It turns out that at LO 3 extra terms appear in the chiral Lagrangian as was discussed in \cite{Sharpe} (note the different sign convention)\footnote{One may wonder how it is possible to describe QCD in the Wilson discretisation scheme by a chiral Lagrangian in the continuum. 
Close to the continuum limit this is a good approximation, see \cite{Sharpe} for further details.}. They contain 3 new low energy constants (LEC) $W_6$, $W_7$ and $W_8$, and all terms are proportionally to $a^2$:
\begin{eqnarray}
{\cal Z}_{\rm WechPT}^{[N_f]}(\{\hat{m}\})&=&
\int_{U(N_f)} dU_0\det[U_0]^\nu\exp\Big[+\frac12 \Sigma V\mbox{Tr}({M}(U_0+U_0^\dag))
-a^2V
W_8\mbox{Tr}(U_0^2+U_0^{\dag\,2})\nonumber\\
&&\quad\quad\quad\quad\quad\quad\quad\quad\quad\quad\quad
-a^2 VW_6(\mbox{Tr}(U_0+U_0^\dag))^2-a^2 VW_7(\mbox{Tr}(U_0-U_0^\dag))^2
\Big]\ .
\label{ZWechPT}
\end{eqnarray}
It is clear that this implies the scaling $\hat{a}^2=a^2V$ or $a\sim \epsilon^{2}$, in addition to the scaling of the quark masses with the volume in eq. (\ref{exponenteps}). 
Let us discuss some simplifications first. Without fixing topology in the special case of $SU(2)$ the term Tr$(U_0-U_0^\dag)$ vanishes, and the terms proportional to $W_8$ and $W_6$ are equivalent. In general, in the second line of eq. (\ref{ZWechPT}) the 2 squares of the traces can be linearised by so-called Hubbart-Stratonovich transformations, at the expense of 2 extra Gaussian integrals. The linearised terms can be included into a shift of the mass for $W_6$, and into so-called axial mass terms $\frac12\mbox{Tr}(\hat{Z}(U_0-U_0^\dag))$ for $W_7$, see \cite{ADSV} for more details. Also for the latter term the corresponding group integral generalising eq. (\ref{LSint}) is known, see \cite{SchlittgenWettig}. Therefore in the following we will set $W_6=0=W_7$, having in mind that we need to perform 2 extra integrals for the full partition function (\ref{ZWechPT}) for non-zero values of these LECs.

Let us define $\hat{a}_8^2=a^2 V W_8$ and compute the partition function only including this term. For $N_f=1$ we obtain
\begin{equation}
{\cal Z}_{\rm WechPT}^{[N_f=1]}(\hat{m})=\int_{-\pi}^\pi\frac{d\theta}{2\pi} e^{i\nu\theta+\hat{m}\cos(\theta)-4\hat{a}_8^2\cos(\theta)^2+2\hat{a}_8^2} 
=e^{2\hat{a}_8^2}\int_{-\infty}^\infty \frac{dx}{\sqrt{\pi}}e^{-x^2}\left(\frac{\hat{m}-4ix\hat{a}_8}{\hat{m}+4ix\hat{a}_8}\right)^{-\frac{\nu}{2}}
I_\nu\left(\sqrt{\hat{m}^2+16x^2\hat{a}_8^2}\right).
\label{ZWechPTNf1}
\end{equation}
It is no longer elementary, but after linearising the $\cos(\theta)^2$ term it can be written as a Gaussian integral over the one-flavour partition function $I_\nu(\hat{m})$ in the continuum at shifted mass. For degenerate masses and general $N_f$ we can again diagonalise $U_0$ and use Andr\'eief's integral formula, as in the derivation of eq. (\ref{Zdeg}), leading to:
\begin{equation}
{\cal Z}_{\rm WechPT}^{[N_f]}(\hat{m})=\det_{1\leq j,k\leq N_f}\left[e^{2\hat{a}_8^2} \int_{-\pi}^\pi\frac{d\theta}{2\pi}\exp\left[i(\nu+k-j)\theta+\hat{m}\cos(\theta)-4\hat{a}_8^2\cos(\theta)^2\right]\right].
\label{ZWechPTNf}
\end{equation}
Once again it is given by the determinant of $N_f=1$ flavour partition functions. In order to determine the spectral density of $D_5$ or $D_W$ one then has to introduce additional auxiliary quark pairs as described in eq. (\ref{QCDresolvent}) (or use replicas). Prior to the RMT calculation this was done for the spectral density from WechPT in \cite{DSV,ADSV}, for both 
$D_5$ or $D_W$.

Let us now turn to the RMT side as introduced in \cite{DSV}. For simplicity we will only consider the Hermitian operator $D_5$, for the discussion of the complex eigenvalue spectrum of $D_W$ we refer to \cite{ADSV} and to \cite{MK} for its complete solution.
Actually two slightly different RMT have been proposed, which have turned out to be equivalent on the level of the jpdf. In \cite{DSV} the partition function (\ref{matrixrep}) was generalised using eq. (\ref{DWchiral}) in an obvious way, by adding two Hermitian Gaussian matrices $A$ and $B$ of dimensions $N$ and $N+\nu$, respectively: 
\begin{equation}
Z_{WI,\,N}^{[N_f]}\sim\int[dW][dA][dB]\prod_{f=1}^{N_f}\det
\left[
\begin{array}{cc}
m_f 1_N+aA&i W\\
-iW^\dag &-m_f 1_{N+\nu}-aB\\
\end{array}
\right] \exp\left[-\frac14\mbox{Tr}(A^2+B^2+2WW^\dag)\right].
\label{WImatrixrep}
\end{equation}
In \cite{AN} the following modification was made: rather than adding a block-diagonal Hermitian matrix to $D_5$ as the last term in eq. (\ref{DWchiral}), a full Hermitian matrix $H$ was added to it, $D_5=\gamma_5(D\!\!\!\!/\ +m)+H$, filling the off-diagonal blocks with a rectangular complex matrix $\Omega$ (and trivially changing the sign of $B\to-B$):
\begin{eqnarray}
Z_{WI\!I,\,N}^{[N_f]}&=&\int[dW][dH]\prod_{f=1}^{N_f}\det
\left[
\begin{array}{cc}
m_f 1_N+A&i W+\Omega\\
-iW^\dag +\Omega^\dag&-m_f 1_{N+\nu}+B\\
\end{array}
\right] \exp\left[-\frac{1}{2(1-a^2)}\mbox{Tr}WW^\dag-\frac{1}{4a^2}\mbox{Tr}H^2\right],\nonumber\\
&& H=\left(\begin{array}{cc}
A&\Omega\\
\Omega^\dag&B\\
\end{array}\right),\ \ a\in[0,1]\ .
\label{WIImatrixrep}
\end{eqnarray}
The different rescaling of the variances with $a$ was made in \cite{AN} to underline that on the level of RMT this corresponds to a two-matrix model that interpolates between the chGUE and the GUE, in the limits $a\to0$ and $a\to1$, respectively. It was shown in \cite{DSV,ADSV} and \cite{AN} that in the limit $N\to\infty$, identifying $\hat{a}_8^2=a^2N/4$ in addition to the rescaled mass terms the two RMT partition functions and eq. (\ref{ZWechPT}) with $W_6=W_7=0$ agree:
\begin{equation}
{\cal Z}_{\rm WechPT}^{[N_f]}=\lim_{\stackrel{\mbox{\small $N\to\infty$}}{\mbox{\small $a,m_f\to0$}}} {\cal Z}_{WI,\,N}^{[N_f]}=
\lim_{\stackrel{\mbox{\small $N\to\infty$}}{\mbox{\small $a,m_f\to0$}}} {\cal Z}_{WI\!I,\,N}^{[N_f]}\ .
\label{Wequiv}
\end{equation} 
As we have said before the presence of the $W_6$- and $W_7$-terms can be achieved by 2 extra Gaussian integrals, see \cite{ADSV}.
It is quite remarkable that RMT allows to include such a detailed structure of WchPT.
The fact that we have ``only'' considered RMTs for $D_5$ does not affect the equivalence argument of the partition functions. Because of eq. (\ref{D5}) and $\det[\gamma_5]=1$ the partition functions of $D_W$ and $D_5$ agree. 

Let us just state the jpdf of the eigenvalues $d_j$ of $D_5$ for degenerate masses $m$ without derivation, see \cite{AN} and \cite{MK} for details,
\begin{eqnarray}
{Z}_{W,\,N}^{[N_f]}(m) &\sim&
\exp\Big[\frac{Nm^{2}}{2a^2(1-a^2)}\Big]
\int_{-\infty}^\infty\prod_{j=1}^{2n+\nu}
dd_jd_j^{N_f}
\exp\Big[{-\frac{d_j^2}{4a^2}}\Big]
\Delta_{2n+\nu}(\{d\})\ \ \ \ 
\nonumber\\
&&\times\mbox{Pf}_{1\leq i,j\leq 2n+\nu;\,1\leq q\leq \nu}
\left[
\begin{array}{cc}
F(d_j-d_i)& \!d_i^{\, q-1}\exp\left[-\frac{d_im}{2a^2}\right]\\
 -d_j^{\, q-1}\exp\left[-\frac{d_jm}{2a^2}\right] & {0}\\
\end{array}
\right]\ ,
\label{ZDevPf}
\end{eqnarray}
where we have defined the antisymmetric weight
\begin{equation}
\label{Fdef}
F(x)= \frac{4}{\sqrt{2\pi a^2(1-a^2)}}\int_m^\infty du\
\exp\left[-\frac{u^2}{2a^2(1-a^2)}\right] 
\sinh\Big[\frac{xu}{2a^2}\Big].
\end{equation}
This jpdf has two new features. First, there appears a Vandermonde times a Pfaffian (Pf) defined as $\mbox{Pf}[A]=\sqrt{\det[A]}$ for antisymmetric matrices $A$ of even dimension, see e.g. \cite{Mehta} for a definition in terms of permutations. This is quite common for such interpolating two-matrix models, see \cite{MehtaPandey} for classical works regarding the GUE-GOE and the GUE-GSE transitions. Second, and most importantly the solution for the correlation functions involves both skew OP and OP at the same time, where the OP are Hermite polynomials of a GUE of size $\nu\times\nu$.  We refer to \cite{MK} for details, in particular that this sub-structure of a finite size GUE remains valid in the large-$N$ limit. In \cite{AN} where only the cases $\nu=0$ and 1 were solved this structure was not observed, because only Hermite polynomials of degree 0 and 1 occur.

\begin{figure}[h]
\unitlength1cm
\begin{center}
\begin{picture}(12.2,5.5)
\includegraphics[width=7cm]{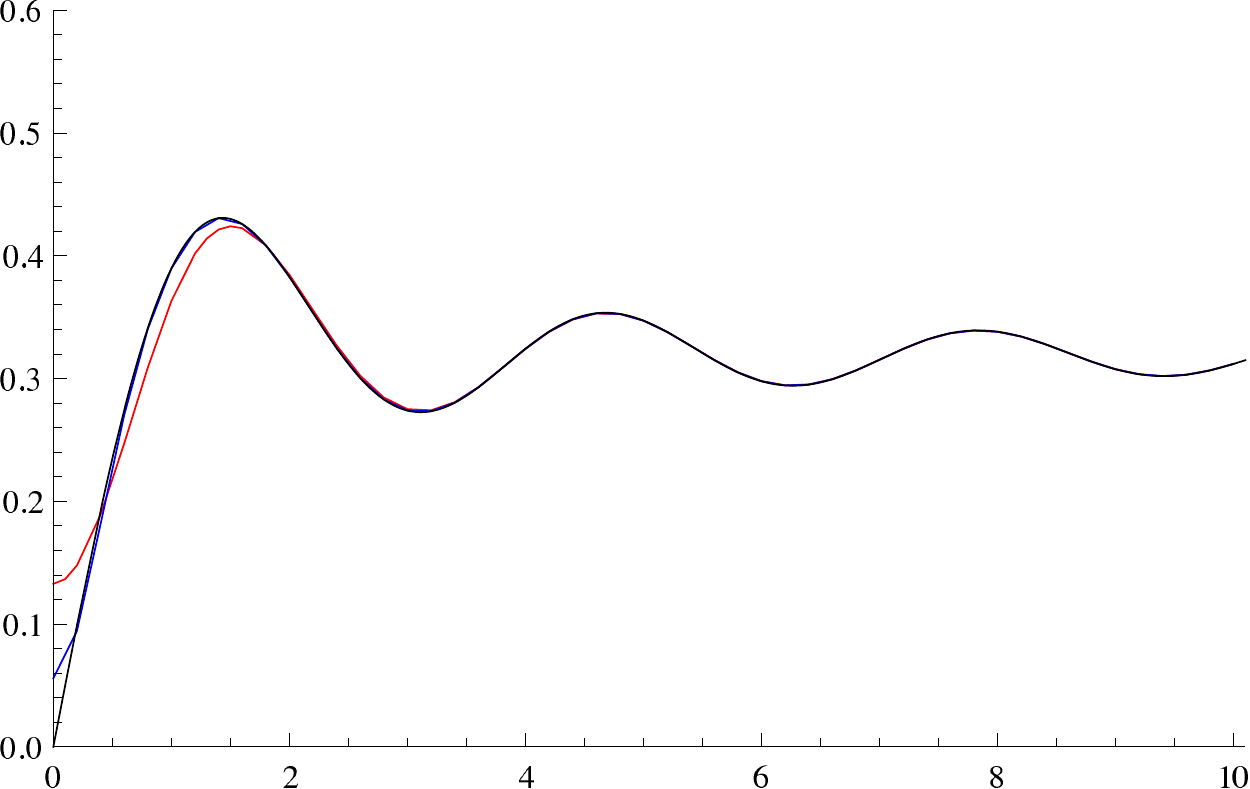}
\includegraphics[width=7cm]{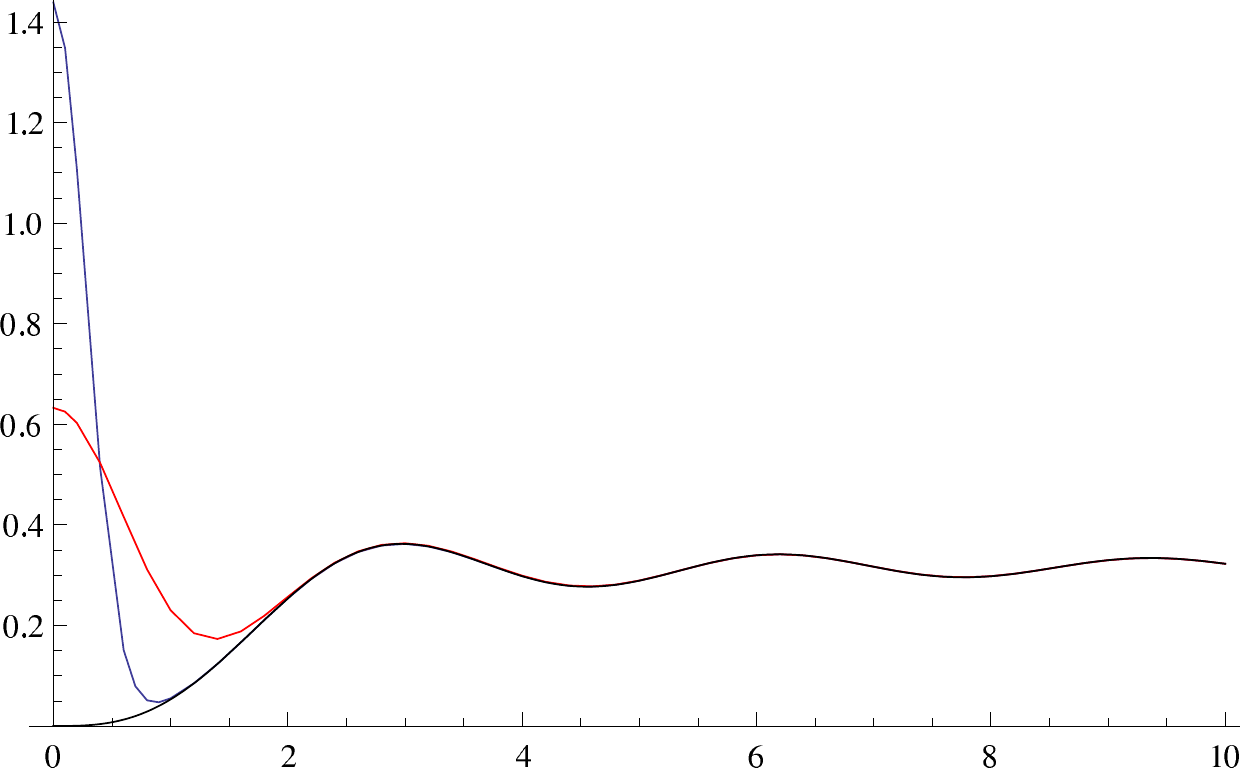}
\put(0.1,-0.2){$\tilde{x}$}
\put(-7.1,-0.2){$\tilde{x}$}
\put(-10.8,3.5){$\rho_s^{D_5}(\tilde{x})$}
\put(-3.5,1.4){$\rho_s^{D_5}(\tilde{x})$}
\end{picture}
\end{center}
\caption{The quenched microscopic spectral density $\rho_s^{D_5}(\tilde{x})$ for $\nu=0$ (left) and $\nu=1$ (right)  as a function of $\hat{a}_8$, taken from \cite{AN}. Because of choosing $\hat{m}=0$ it is still symmetric around the origin. For $\hat{a}_8=0$ we are back to the microscopic density of Bessel functions as in Figure \ref{fig:Besseldensity}, whereas for $\hat{a}_8=0.1$ (blue) and $0.25$ (red) the density starts to spread into the region around the origin. For $\nu=1$ the delta-function of the one zero-eigenvalue is broadened by increasing $\hat{a}_8$ - which is why this is plotted on a different scale compared to $\nu=0$.  At these small values of $\hat{a}_8$ the effect on the density is remarkably localised to the origin. 
Only in the limit $\hat{a}_8\to\infty$ we will arrive at the microscopic density of the GUE \cite{AN} which is completely flat (and normalised to $1/\pi$ here).
\label{fig:WBessel}}
\end{figure}

Instead of giving any further details of the solution of WRMT we show plots in Figure \ref{fig:WBessel} below for the limiting quenched microscopic density $\rho_s^{D_5}(\tilde{x})$ of $D_5$ for $\nu=0,1$, illustrating the effect of the rescaled finite lattice spacing $\hat{a}$. In order to compare to the results from the chGUE in Figure \ref{fig:Besseldensity} let us first discuss the difference between $D_5$ and $D\!\!\!\!/$ at $a=0$,
where  $D_5=\gamma_5(D\!\!\!\!/+m)$,  see eqs. (\ref{WDirac}) and (\ref{D5}). The rotation with $\gamma_5$ merely makes the Dirac operator Hermitian rather than anti-Hermitian, which corresponds to rotating the eigenvalues from the imaginary to the real axis (as we had already done in writing the jpdf in eq. (\ref{partitionfunct}) of real eigenvalues). 
The shift by $\gamma_5 m$ however introduces a gap such that there are no eigenvalues of $D_5$ in $[-{m},{m}]$. Furthermore, the  $\nu$ exact zero eigenvalues which are not shown in Figure \ref{fig:Besseldensity} get moved away from the origin to one of the edges of this gap, which in our convention is at $-{m}$. Therefore we now have to give the density on $\mathbf{R}$ instead of $\mathbf{R}_+$, and it holds for $N_f=0$:
\begin{equation}
\lim_{\hat{a}_8\to0}\rho_s^{D_5}(\tilde{x})=\frac{ |\tilde{x}|}{2}\Theta(\tilde{x}-\tilde{m})\left(J_\nu(\sqrt{\tilde{x}^2-\tilde{m}^2})^2- 
J_{\nu-1}(\sqrt{\tilde{x}^2-\tilde{m}^2})J_{\nu+1}(\sqrt{\tilde{x}^2-\tilde{m}^2}) 
\right) +\nu\delta(\tilde{x}+\tilde{m})\ .
\label{rhoD5a0}
\end{equation}
The effect of $\hat{a}_8>0$ is both to broaden the $\nu$ delta-functions and to eventually fill the gap $[-\hat{m},\hat{m}]$. Because eigenvalues repel each other the $\nu$ zero-modes will now spread, and form approximately a GUE of finite size $\nu$, see \cite{ADSV} for a further discussion. In order to make the comparison more transparent let us also choose $\hat{m}=0$, so that we can plot the density on $\mathbf{R}_+$ only, as is shown in Figure \ref{fig:WBessel}. The effect of $\hat{a}_8>0$ is particularly striking on the zero-modes.

The density of the Hermitian Wilson Dirac operator $D_5$ has been compared to quenched lattice simulations in \cite{PUK,Wengeretal}. Because this involves multi-parameter fits several quantities have been proposed that are simple to measure \cite{KVZ}. A comparison to unquenched data remains a difficult question, see however \cite{Cichy}.
The RMT approach presented here has also lead to insights about the sign and constraints amongst the LECs $W_{6,7,8}$ that were previously controversial \cite{ADSV,ADSVII}.  When the lattice artefacts become very strong additional unphysical phases appear, and the possibility to access these as a function of the LEC and quark content $N_f$ was clarified in \cite{KVZ2}. 
The derivation of all $k$-point correlation functions including $N_f\neq0$ based on WechPT has been pushed forward in \cite{KimJac2011}, leading to alternative representations compared to the OP approach \cite{MK}. For the density of $D_5$ in the $p$-regime see \cite{Necco}.
The computation of individual eigenvalue distribution is not known beyond the expansion of the Fredholm expansion proposed in \cite{AIp1}.

\subsection{RMT and QCD with chemical potential}\label{RMTmu}

In this subsection we will study the influence of a chemical potential $\mu$ on the Dirac operator spectrum. We will first state its global symmetry in QCD, then construct the RMT and at the end link to the epsilon-regime of the corresponding chiral Lagrangian. The addition of a chemical potential $\mu_q$ for each quark flavour to the QCD action in eq. (\ref{ZQCD}) amounts to add the terms $\mu_q\Psi^{q\,\dag}(x)\Psi^q(x)=\overline{\Psi}^q(x)\mu_q\gamma_4\Psi^q(x)$. In classical statistical mechanics the addition of a chemical potential allows the number of the corresponding particles to fluctuate.  Here we will add the same chemical potential $\mu_q=\mu$ to all flavours (which makes it the Baryon chemical potential).
In analogy to eq. (\ref{Dchiral}) the global symmetry of the Dirac operator plus $\mu\gamma_4$ changes to 
\begin{equation}
D\!\!\!\!/(\mu)=
\left( 
\begin{array}{cc}
0&i{\cal W}+\mu\\
i{\cal W}^\dag+\mu&0\\
\end{array}
\right)\neq -D\!\!\!\!/(\mu)^\dag\ ,
\label{Dchiralmu}
\end{equation}
and thus becomes complex non-Hermitian. It holds that $-D\!\!\!\!/(-\mu)^\dag=D\!\!\!\!/(\mu)$, and thus for purely imaginary chemical potential $\mu\to i\mu$ it is still anti-Hermitian. 
Because the operator remains chiral
\begin{equation}
0=\{D\!\!\!\!/(\mu),\gamma_5\}\ ,\ \   \ D\!\!\!\!/(\mu)\Phi_k(x)=iz_k\Phi_k(x)\ ,
\label{Dmuchiral}
\end{equation}
the non-zero complex eigenvalues continue to come in pairs $\pm iz_k$, using the same argument as after eq. (\ref{eigenfunct}).
The fact that the spectrum of the Dirac operator with chemical potential is complex (and thus the QCD action too) has severe repercussions for lattice QCD. Because the action can no longer be interpreted as a probability weight, the standard method on the lattice, weighting a given configuration by its action breaks down. Several methods have been invented to circumvent this so-called sign problem, and we refer to \cite{PdF} for a review. In contrast the RMT for QCD with chemical potential remains analytically solvable with OP in the complex plane, as we will see. 

In \cite{Mischa} a RMT was constructed by Stephanov where ${\cal W}$ was replaced by a complex Gaussian random matrix $W_1$, cf. eq. (\ref{matrixrep}) for $\beta=2$ compared to (\ref{Dchiral}). While this RMT was very useful to illuminate the difference between $N_f=0$ and $N_f>0$, which is more dramatic than for $\mu=0$, this model was not analytically tractable with OP at finite-$N$. Therefore Osborn \cite{James} introduced a two-matrix model where $\mu$ is multiplied by a second random matrix $W_2$, see eq. (\ref{2matrixrep}) below, implying that the chemical potential term is not diagonal in that basis. While at first sight more complicated this model turns out to be exactly solvable using OP in the complex plane, as we will demonstrate. 
A further equivalent representation was chosen in \cite{Blochetal}, multiplying $W_{1,2}$ by $\exp[\mp\mu]$ instead. This parametrisation allowed to understand, why the sign problem in the numerical solution of this RMT for QCD with $\mu\neq0$ can be circumvented \cite{Jacsubset}.
Prior to introducing an RMT with $\mu$ as in eq. (\ref{Dchiralmu}) a similar model was introduced \cite{MischaT} describing the effect of temperature $T$. 
Such a model can only be solved using the supersymmetric method, see \cite{Seif}, with the effect of renormalising $\Sigma\to\Sigma(T)$ to be temperature dependent, keeping the correlation functions otherwise unchanged.
A further modification of Stephanov's model by including both parameters $\mu\to \mu+i\pi T$ in eq. (\ref{Dchiralmu}) was very successful as a schematic model for the phase diagram of QCD \cite{Halasz}, predicting the qualitative features depicted  in Figure \ref{fig:phasediag}.

Let us turn to the RMT \cite{James} defined as 
\begin{eqnarray}
Z_N(\mu)&=&\int[dW_1][dW_2]\prod_{f=1}^{N_f}\det
\left[
\begin{array}{cc}
m_f 1_N&i W_1+\mu W_2\\
iW_1^\dag +\mu W_2^\dag &m_f 1_{N+\nu}\\
\end{array}
\right] \exp[-\mbox{Tr}(W_1W_1^\dag +W_2W_2^\dag)]\nonumber\\
&\sim& \int[dX_1][dX_2]\prod_{f=1}^{N_f}\det
\left[
\begin{array}{cc}
m_f 1_N&X_1\\
X_2&m_f 1_{N+\nu}\\
\end{array}
\right] \exp\left[-a(\mu)\mbox{Tr}(X_1X_1^\dag +X_2X_2^\dag)\right]\nonumber\\
&&\quad\quad\quad
\times\exp\left[b(\mu)\mbox{Tr}(X_1X_2+X_2^\dag X_1^\dag)\right]\ , \ \ a(\mu)=\frac{1+\mu^2}{4\mu^2}\ , \ \ b(\mu)=\frac{1-\mu^2}{4\mu^2}\ .
\label{2matrixrep}
\end{eqnarray}
In the second step we have simply changed variables to the matrices  $X_1=iW_1+\mu W_2$ and $X_2=iW^\dag+\mu W_2^\dag$ which are now coupled. While at $\mu=1$ the matrices $X_1$ and $X_2$ are again independent Gaussian random matrices - this is called maximal non-Hermiticity - at $\mu=0$ we have $X_1^\dag=-X_2$ with total correlation (and we are back to the chGUE). A look at the characteristic equation for the massless RMT Dirac operator, 
\begin{equation}
\det[z-D\!\!\!\!/(\mu)]=\det[z^2-X_1X_2]\ ,
\label{muchar}
\end{equation}
reveals that we are after the complex eigenvalues\footnote{For the singular values of $X_1X_2$ in this setting see \cite{AStr}.} of the product of the two correlated matrices $Y=X_1X_2$.
As in eq. (\ref{matrixrep}) we could have allowed for real or quaternion valued matrix elements with $\beta=1,4$. However, the joint density of complex eigenvalues does not enjoy a closed form for all 3 $\beta$'s as in eq. (\ref{partitionfunct}). In particular real matrices are special as they can have real eigenvalues and complex conjugated eigenvalues pairs, as the characteristic equation (\ref{muchar}) is real. For the solution of the respective RMTs see \cite{APSo} and \cite{A05}, respectively.

The parametrisation of a generic  complex non-Hermitian matrix $Y$ in terms of its complex eigenvalues is usually done using the Schur decomposition $Y=U(Z+T)U^\dag$, where $U\in U(N)$ is unitary, $Z=\mbox{diag}(z_1,\ldots,z_N)$ contains the complex eigenvalues, and $T$ is a strictly upper triangular complex matrix (it can be chosen to be strictly lower triangular instead). For two matrices however, in order to compute the Jacobian we have to use the following generalised Schur decomposition \cite{Golub}
\begin{equation}
X_1=U(Z_1+R)V\ ,\ X_2=V^\dag(Z_2+S)U^\dag\ \Rightarrow\ \ Y=X_1X_2=U(Z+T)U^\dag\ ,
\label{Schur}
\end{equation}
with $Z=Z_1Z_2$ and $T=RZ_2+Z_1S+RS$.
Here $U,V\in U(N)$ are unitary, $Z_1,Z_2$ are diagonal matrices with complex entries (which are not the complex eigenvalues of $X_1$ and $X_2$), and $R,S$ are strictly upper triangular. The complex eigenvalues of the product matrix $Y$ are thus given by the elements of the diagonal matrix $Z$. Let us directly give the result for the joint density from \cite{James} (see \cite{KanzieperSingh} for an alternative derivation):
\begin{eqnarray}
Z_N(\mu)^{[N_f]}&=&
\left(\prod_{j=1}^N\int d^2z_j|z_j|^{\nu}K_\nu(a(\mu)|z_j|)\exp\left[-\frac{b(\mu)}{2}(z_j+z_j^*)\right]\prod_{f=1}^{N_f}(z_j+m_f^2)\right) |\Delta_N(\{z\})|^2
\label{ZRMTmu}
\\
&=& \left(\prod_{j=1}^N\int d^2z_j\right) {\cal P}_{jpdf}(z_1,\ldots,z_N)\ . 
\label{jpdfmu}
\end{eqnarray}
Note that in view of eq. (\ref{muchar}) we should later take the square root of the eigenvalues $z_j$ of $Y=X_1X_2$ to change to Dirac operator eigenvalues. In the complex plane this is not unique and without loss of generality one can then restrict to the half-plane, also in view of the $\pm$ pairs of Dirac eigenvalues.  
Before we discuss the properties of eq. (\ref{ZRMTmu}) and its correlation functions let us try to understand why the Jacobian of the transformation (\ref{Schur}) does not only include the modulus squared Vandermonde determinant (as in the complex Ginibre ensemble, see \cite{Mehta}) and the zero-modes $|z_j|^\nu$, but also a modified $K$-Bessel function, even though we started from (coupled) Gaussian matrices. The same phenomenon happens when looking at products of complex Gaussian random variables $z_{1,2}$, and asking for the distribution $w(z)$ of the product random variable $z=z_1z_2$:
\begin{equation}
w_2(z)=\int d^2z_1 \int d^2z_2 \exp[-|z_1|^2-|z_1^2|]\delta^{(2)}(z-z_1z_2)=(2\pi)^2 \int_0^\infty \frac{drr}{r^2}\exp[-r^2-|z|^2/r^2]= 2\pi K_0(2|z|)\ .
\label{Kweight}
\end{equation}
Obviously we are dealing with the analogue of quadratic matrices here.
Multiplying $n\geq2$ such random variables we obtain an $(n-1)$-fold integral representation for the distribution of $z$ now given by a Meijer G-function:
\begin{equation}
w_n(z)=\prod_{j=1}^n\int d^2z_j e^{-|z_j|^2}\delta^{(2)}(z-z_1\cdots z_n)=
(2\pi)^{n-1}
\prod_{j=1}^{n-1}\int_0^\infty \frac{dr_j}{r_j} e^{-r_j^2}\ 
e^{-\frac{|z|^2}{r^2_1\cdots r^2_{n-1}}}
=\pi^{n-1}G^{n\,0}_{0\,n}\left(\mbox{}_{\vec{0}}^{-} \bigg| \, |z|^2
\right) ,
\end{equation}
see \cite{GradshteynRyzhik} for its standard definition.
These weights appear when studying the complex eigenvalues of the product of $n$ independent Gaussian matrices $X_1,\ldots,X_n$ and we refer to \cite{ABu} for the solution of this model.
 
The complex eigenvalue correlation functions of the model (\ref{ZRMTmu}) are defined as in eq. (\ref{Rk}) and it holds 
\begin{equation}
R_k(z_1,\ldots,z_k)=\prod_{j=1}^kw(z_j)\det_{1\leq i,j\leq k}[K_N(z_i,z^*_j)]\ .
\label{RkC}
\end{equation}
They are again given in terms of the kernel
\begin{equation}
K_N(z,u^*)=\sum_{j=0}^{N-1} h_j^{-1}P_{j}(z)Q_j(u^*)
\label{mukernel}
\end{equation}
of polynomials $P_j(z)$ and $Q_j(z)$ which are bi-orthogonal in the complex plane with respect to the weight 
\begin{eqnarray}
w(z)&=&|z_j|^{\nu}K_\nu(a(\mu)|z_j|)\exp\left[-\frac{b(\mu)}{2}(z_j+z_j^*)\right]\prod_{f=1}^{N_f}(z_j+m_f^2)\ , 
\label{muweight}\\
&&\ \ \int d^2z\ w(z) P_k(z)Q_l(z)=h_k\delta_{kl}\ .
\label{OPC}
\end{eqnarray}
We will not discuss gap probabilities or individual eigenvalue distributions here and refer to \cite{APS}. Unlike in the real eigenvalue case they will now depend on the choice of the geometry of the region containing $k\geq0$ eigenvalues, and only in the rotationally invariant case $\mu=1$ a radial ordering of eigenvalues according to their modulus provides a natural choice.

Before we solve the RMT eq. (\ref{ZRMTmu}) a word of caution is in order. In standard OP theory in the complex plane only real positive weights are considered, which is only true for $N_f=0$ in eq. (\ref{muweight}). In this case $P_k(z)^*=Q_k(z)$ and standard techniques as Gram-Schmidt apply to construct these polynomials, see e.g. \cite{Walter}. When $N_f>0$ in our case the weight is complex, and thus a priori the partition function is as well as the $R_k$ are. Therefore the density no longer has a probabilistic interpretation, it has a real and imaginary part, and we refer to \cite{AOSV} for a more detailed discussion.

We will first provide the solution of the simplest quenched case with $N_f=0$.  Then the partition function and general eigenvalue correlation functions including $N_f$ mass terms can be obtained along the same lines as in Section \ref{OPDirac}, expressing them in terms of the OP at $N_f=0$. 
Surprisingly, despite the complicated non-Gaussian weight function in the complex plane the OP are again given by Laguerre polynomials of complex arguments \cite{James}
\footnote{A first unsuccessful attempt with the incorrect weight (except for $\nu=\pm\frac12$) was made in \cite{A02}.}. The following holds, where we drop the argument $\mu$ of the parameters $a>b\geq0$ in eq. (\ref{2matrixrep}):
\begin{equation}
\int d^2 z|z|^\nu K_\nu(a|z|)\exp[b\Re e(z)] L_j^\nu(cz)L_k^\nu(cz)=\delta_{jk}h_k\ ,\ \mbox{with} \ \ c=\frac{a^2-b^2}{2b}\ ,\ \ h_j=\frac{\pi(j+\nu)!}{j!\,a}\left(\frac{a}{b}\right)^{2j}\left(\frac{ac}{b}\right)^{\nu+1}\ .
\label{OPmu}
\end{equation}
Here $\nu=0,1,\ldots$ is an integer, and the Laguerre polynomials can be made monic as in eq. (\ref{Laguerre}). The proof we sketch follows \cite{ABender} and uses induction of depth 2, for an earlier longer proof see \cite{A05}. Insert the following integral representations into the orthogonality relation (\ref{OPmu}) for $\nu=0,1$,
\begin{eqnarray}
K_\nu(x)&=&\frac{x^\nu}{2^{\nu+1}}\int_0^\infty \frac{dt}{t^{\nu+1}}\exp[-t-x^2/(4t)]\ ,\nonumber\\
L_j^{\nu}(z)&=&\oint_\Gamma \frac{du}{2\pi} \frac{\exp\left[-\frac{zu}{1-u}\right]}{(1-u)^{\nu+1}u^{j+1}}\ ,
\label{cintrep}
\end{eqnarray}
where $\Gamma$ is a contour enclosing the origin in positive direction, but not the point $z=1$. Then the Gaussian integrals over the real and imaginary parts of $z=x+iy$ can be solved, and subsequently the integral over $t$ from the $K$-Bessel function. The orthogonality then easily follows applying the residue theorem twice. For the induction we use the 3-step recurrence relation for the Laguerre polynomials and $K$-Bessel function \cite{GradshteynRyzhik}, see \cite{ABender} for details. We thus have solved the quenched case as we know all correlation functions from the kernel
\begin{equation}
K_N(z,u^*)=\sum_{j=0}^{N-1} h_j^{-1}L_j^{\nu}(cz)L_j^{\nu}(cu^*)\ ,
\label{cLagKernel}
\end{equation}
in terms of the quantities in eq. (\ref{OPmu}), by using eq. (\ref{RkC}). A similar solution exists \cite{FKS} for the so-called elliptic complex Ginibre ensemble of a single complex matrix $W=H+iA$, 
where the Hermitian part $H=H^\dag$ and anti-Hermitian part $A=A^\dag$ of the matrix have different variances that depend on a parameter. The solution is given in term of Hermite polynomials in the complex plane and allows to interpolate between the GUE and the complex Ginibre ensemble at maximal non-Hermiticity, where $H$ and $A$ have the same variance. We refer to \cite{FKS} for a detailed discussion.

Let us now turn to the unquenched solution with $N_f>0$ described in great detail in \cite{AOSV}, expressing everything in terms of the Laguerre polynomials of the quenched solution from eq. (\ref{OPmu}). Because we follow the same strategy as in Subsections \ref{genweight} and \ref{allLaguerre} using expectation values of characteristic polynomials we can be very brief. The corresponding proofs for manipulating expectation values of characteristic polynomials and their complex conjugates defined as in eq. (\ref{ave}) can be found in \cite{AV03}.
In particular for the monic OP in the complex plane satisfying eq. (\ref{OPC}) for a weight $w(z)$ that can be complex we have \cite{AV03}
\begin{equation}
P_L(z)=\left\langle \prod_{j=1}^L(z-z_j)\right\rangle_L\ ,\ \ Q_L(u^*)=\left\langle \prod_{j=1}^L(u^*-z_j^*)\right\rangle_L\ .
\label{OPCave}
\end{equation}
The proof goes exactly the same way as in eq. (\ref{proofPN}), where for $P_L(z)$ we multiply the product into $\Delta_L(\{z\})$, increasing it to $\Delta_{L+1}$. For $Q_L(u^*)$ we simply choose $\Delta_L(\{z\})^*$, increasing it to $\Delta_{L+1}^*$. Likewise for the kernel we obtain
\begin{equation}
K_{N+1}(z,u^*)=h_N^{-1}\left\langle \prod_{j=1}^N(z-z_j)(u^*-z_j^*)\right\rangle_N=\sum_{j=0}^Nh_j^{-1}P_j(z)Q_j(u^*)\ ,
\label{KernelC}
\end{equation}
modifying the proof in eq. (\ref{proofKernel}) accordingly by multiplying the first product into $\Delta_N(\{z\})$, and the second into $\Delta_N(\{z\})^*$. The only difference here is that in general no Christoffel-Darboux formula is available. This is despite the fact that the Laguerre polynomials in our example eq. (\ref{OPmu}) still satisfy the standard 3-step recurrence (which is not true for OP in the complex plane in general). The step in the proof of eq. (\ref{CD}) that breaks down is the cancellation of the summands after multiplying eq. (\ref{cLagKernel}) by $(z-u^*)$.  

The most general result corresponding to eq. (\ref{prodchar}) now contains both products of $K$ characteristic polynomials and of $L$ products of complex conjugated ones. Without loss of generality let us choose $K\geq L$ ($K<L$ can be obtained by complex conjugation), to state the result from \cite{AV03}:
\begin{equation}
\left\langle \prod_{l=1}^N\left(\prod_{i=1}^K(v_i-z_l)\right)\left(\prod_{j=1}^L(u_j^*-z_l^*)\right)\right\rangle_N=
\frac{\prod_{i=N}^{N+K-1}h_i^{\frac12}\prod_{j=N}^{N+L-1}h_j^{\frac12}}{\Delta_K(\{v\})\Delta_L(\{u\})^*}
\det_{1\leq l,m\leq K}\left[
\begin{array}{cc}
K_{N+L}(v_l,u_m^*)\ \ 
\frac{P_{N+m-1}(v_l)}{h_{N+m-1}^{\frac12}}\\
\end{array}
\right].
\label{Cprodchar}
\end{equation}
Here the left block of kernels inside the determinant is of size $K\times L$. The proof is straightforward and uses induction. Using this result we are now ready to compute the partition function eq. (\ref{ZRMTmu}) for general $N_f\geq0$. In the quenched case it is given again by $N!$ times the products of the norms from eq. (\ref{OPmu}), as derived in eq. (\ref{Znorm}). For $N_f>0$ we can express it in terms of the Laguerre polynomials from eq. (\ref{OPmu}), and obtain an almost identical answer to eq. (\ref{Zmassive}):
\begin{equation}
Z_N^{[N_f]}(\{m\})=
\frac{(-1)^{NN_f}N!\prod_{j=0}^{N-1}h_j}{\Delta_{N_f}(\{m\})}
\det_{1\leq j,g\leq N_f}\left[(-1/c)^{N+g-1}(N+g-1)!L^\nu_{N+g-1}(-cm_f^2)\right].
\label{ZmassiveC}
\end{equation}
As in eq. (\ref{Zmassive}) it is given by a determinant of Laguerre polynomials of real arguments which are real. The only difference to $\mu=0$ is the scale factor $c=1/(1-\mu^2)$ and the $\mu$-dependence inside the norms $h_j$ from eq. (\ref{OPmu}). We will not give further details regarding the kernel 
$K_N^{[N_f]}(z,u^*)$
constituting the $k$-point correlation functions in eq. (\ref{RkC}). Following the first line of eq. (\ref{Kernelmassive}) it can be expressed through the quenched average of $N_f+1$ characteristic polynomials times one conjugated characteristic polynomial, divided by the partition function from eq. (\ref{ZmassiveC}). The spectral density with $N_f$ masses then follows as in eq. (\ref{R1massive}), by multiplying $K_N^{[N_f]}(z,z^*)$ with the weight eq. (\ref{muweight}), see \cite{James,AOSV} for more details. Clearly it is not real in general.

\begin{figure}[h]
\unitlength1cm
\begin{center}
\begin{picture}(12.2,5.5)
\includegraphics[width=7cm]{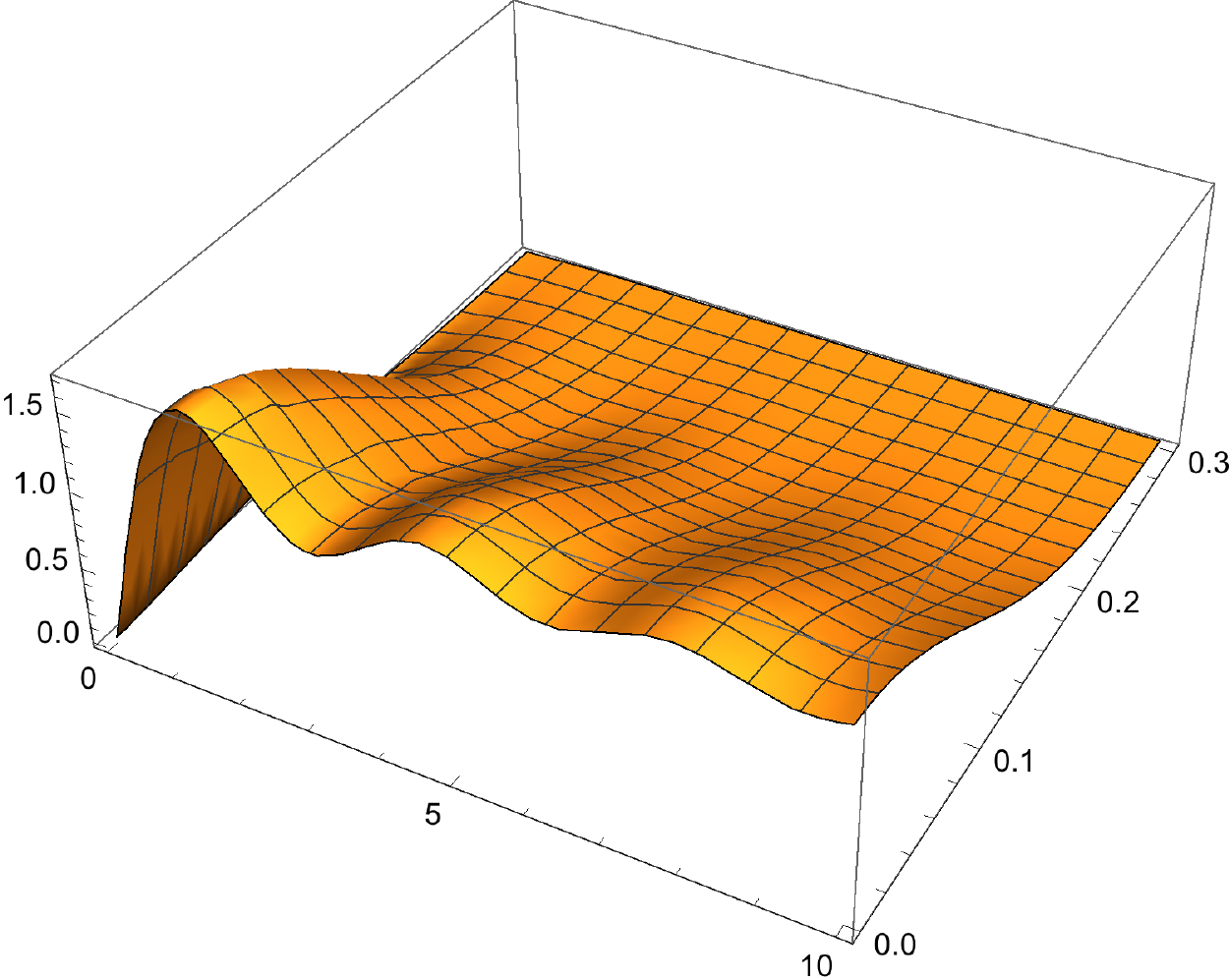}
\includegraphics[width=7cm]{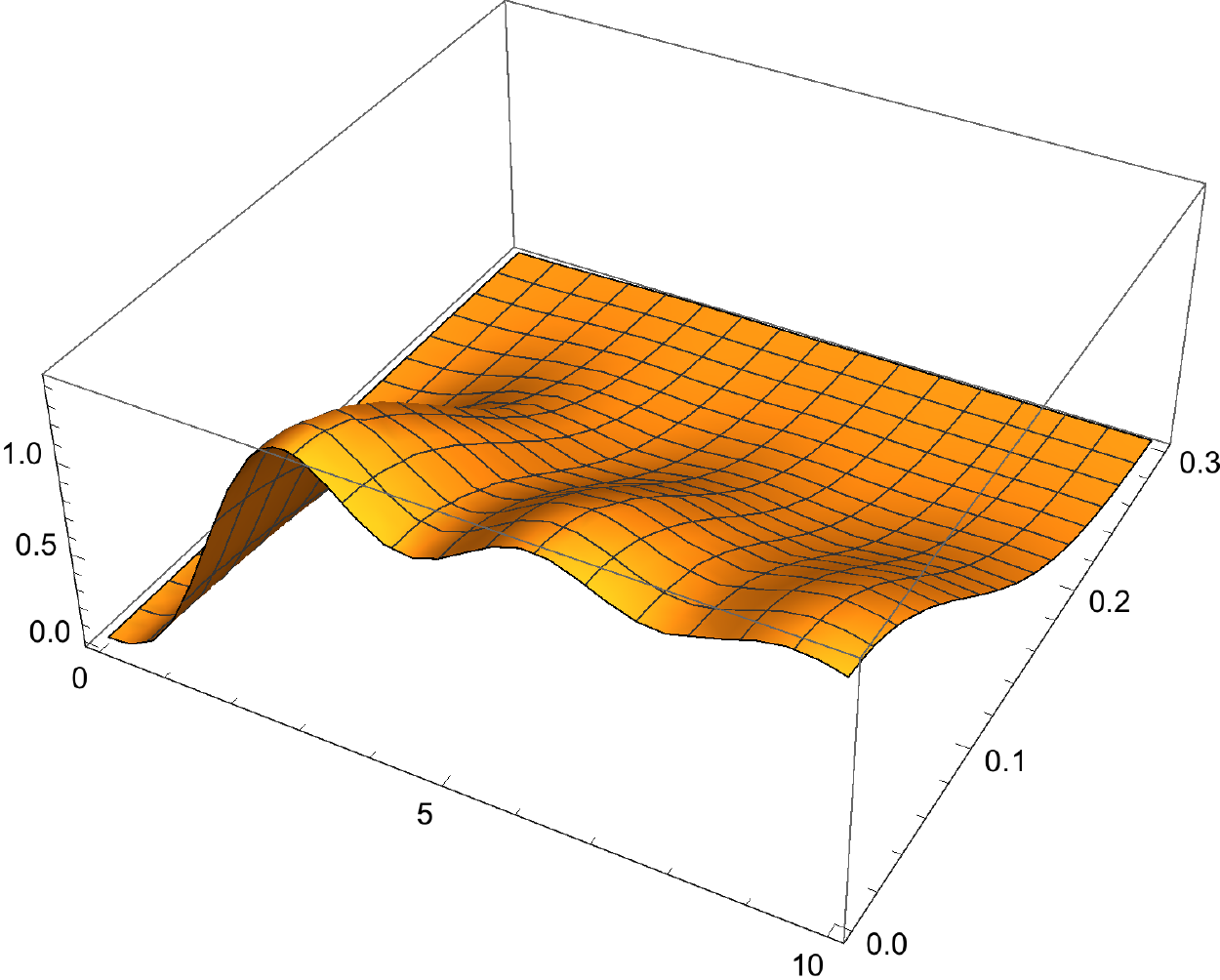}
\put(-5.1,0.8){$\tilde{x}$}
\put(-1.0,1.3){$\tilde{y}$}
\put(-12.1,0.8){$\tilde{x}$}
\put(-8.0,1.3){$\tilde{y}$}
\put(-7.0,4.4){$\rho_s(\tilde{z})$}
\put(-14.0,4.4){$\rho_s(\tilde{z})$}
\end{picture}
\end{center}
\caption{The quenched microscopic spectral density from eq. (\ref{Cdensity}) as a function of $\tilde{z}=\tilde{x}+i\tilde{y}$ for $\nu=0$ (left) and $\nu=1$ (right),  
at $\hat{\mu}=0.1$. The density of real eigenvalues from Figure \ref{fig:Besseldensity} can still be recognised, it now spreads into the complex plane. Because in the limit $\hat{\mu}\to0$ it will become proportionally to a delta-function $\delta(\tilde{y})$ times the Bessel density the vertical scale is different.
\label{fig:BesselC}}
\end{figure}
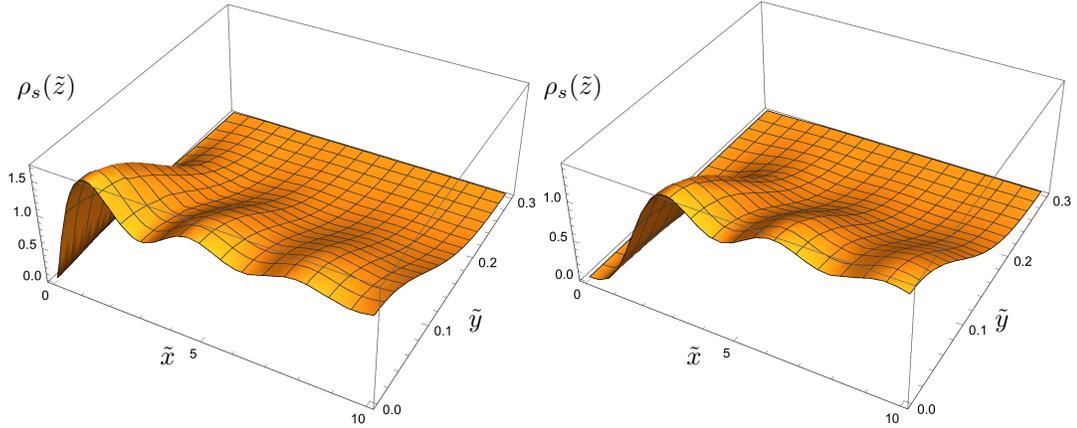

We will now very briefly discuss the large-$N$ limit and matching with the corresponding echPT. It turns out that for non-Hermitian matrices that are coupled as in eq. (\ref{2matrixrep}) the possibility of taking such limits is much richer than in the Hermitian (or maximally non-Hermitian) case. In addition to the standard macroscopic and microscopic statistics a further class of limits exist, where the matrix $Y=X_1X_2$ becomes almost Hermitian, taking the parameter $\mu\to0$ at such a rate that that the following products  
\begin{equation}
\tilde{\mu}^2=\lim_{\stackrel{\mbox{\small $N\to\infty$}}{\mbox{\small $\mu\to0$}}} 2N\mu^2\ ,\ \ \tilde{z}=\lim_{\stackrel{\mbox{\small $N\to\infty$}}{\mbox{\small $z\to0$}}}2\sqrt{N}z
\label{weaklim}
\end{equation}
are constant. Here the scaling of the complex eigenvalues (and masses) is unchanged compared to eq. (\ref{RMTmass}). This limit was first observed in the elliptic complex Ginibre ensemble already mentioned \cite{FKS} and called weakly non-Hermitian \cite{FKSlett}. This is in contrast to the limit at strong non-Hermiticity where $\mu\in(0,1]$ is kept fixed and a different scaling of $z$ is chosen. Here the statistics of the product of two independent complex matrices with $\mu=1$ is found, see \cite{AOSV} and \cite{APS} for details. 
Looking at our result eq. (\ref{ZmassiveC}) it is clear that the limiting partition function is the same as for $\mu=0$ in eq. (\ref{ZRMTNf}) and is thus independent of the rescaled $\hat{\mu}$.  In contrast already the limiting quenched density does depend on $\hat{\mu}$. It can be easily obtained from eq. (\ref{cLagKernel}), replacing the sum by an integral and using eq. (\ref{LaguerreLim}) for the Laguerre polynomials, 
\begin{equation}
\rho_s(\tilde{z})=\frac{|\tilde{z}|^2}{2\pi\hat{\mu}^2}K_\nu(|\tilde{z}|^2/4\hat{\mu}^2)\exp[\Re e(\tilde{z}^2)/4\hat{\mu}^2]\int_0^1 dtt \exp[-2 t^2\hat{\mu}^2]|J_\nu(\tilde{z}t)|^2\ .
\label{Cdensity}
\end{equation}
It is shown in Figure \ref{fig:BesselC}.
Here we have again changed to Dirac eigenvalues following \cite{James}. This result was obtained earlier using replicas and the Toda equation in \cite{KimJacToda}.
For $\hat{\mu}=0$ the integral becomes elementary and matches eq. (\ref{Besseldensity}), while the prefactors turn into a delta-function.

The fact that the partition function is $\hat{\mu}$-independent and the density is not has been called the "Silver blaze problem" \cite{Cohen}, alluding to a novel by A.C. Doyle. 
How can we resolve this puzzle? Let us look again at chPT, the theory of Goldstone Bosons. Because these are Mesons they don't carry Baryon charge and thus do not feel the effect of adding $\mu$ to the action (Mesons consist of a quark and anti-quark whereas Baryons, as the proton, of 3 quarks). We can still make the chPT partition function $\mu$-dependent by adding an integer number $N_f^*$ of quarks with the opposite chemical potential $-\mu$. As discussed after  eq. (\ref{Dchiralmu}) this amounts to add complex conjugated determinants of the Dirac operator to the action, leading to 
\begin{equation}
Z_N^{[N_f+N_f^*]}(\{m\},\{n\})= Z_N \left\langle \prod_{l=1}^N\left(\prod_{i=1}^{N_f}(z_l+m_f)\right)\left(\prod_{j=1}^{N_f^*}(z_l^*+n_f^2)\right)\right\rangle_N\ ,
\label{ZRMTmix}
\end{equation}
which we can also compute using eq. (\ref{Cprodchar}). The corresponding limiting partion function of echPT now does depend on $\mu$ as follows:
\begin{equation}
{\cal Z}_{\rm echPT}^{[N_f+N_f^*]}(\{m\},\{n\})=\int_{U(N_f+N_f^*)} dU_0\det[U_0]^\nu\exp\left[\frac12 \Sigma V\mbox{Tr}({M}(U_0+U_0^\dag))-\frac14\mu^2 VF_\pi^2[U_0,B][U_0,B]\right].
\label{ZechPTmu}
\end{equation}
Here $M=\mbox{diag}(m_1,\ldots,m_{N_f},n_1,\ldots,n_{N_f^*})$ contains both kinds of masses and $B=\mbox{diag}(1_{N_f},-1_{N_f^*})$ is the metric with the signature of the chemical potentials. The commutator comes from the fact that in the Dirac operator $\mu$ acts as an additional vector potential. It obviously vanishes at $N_f^*=0$.
Once again in the limit the partition functions from RMT and echPT agree, when identifying the rescaled masses and chemical potentials as $\hat{\mu}^2=\mu^2 VF_\pi^2=\tilde{\mu}^2$:
\begin{equation}
\lim_{\stackrel{\mbox{\small $N\to\infty$}}{\mbox{\small $m_f,n_f,\mu\to0$}}} {\cal Z}_N^{[N_f+N_f^*]}(\{m\},\{n\})={\cal Z}_{\rm echPT}^{[N_f+N_f^*]}(\{\hat{m}\},\{\hat{n}\}). 
\end{equation}
They are given by the determinant from eq. (\ref{Cprodchar}) in terms of the limiting kernel and the limiting polynomials which are again $I_\nu(\hat{m})$. The corresponding group integral eq. (\ref{ZechPTmu}) was computed independently in \cite{KimJacToda} using an explicit parametrisation of $U(N_f+N_f^*)$, see \cite{AFV} for the general $N_f$ case.

What about the equivalence of RMT and echPT for the density, in other words how can we generate the spectral density eq. (\ref{Cdensity}) from a chiral Lagrangian? Although one may still define and generate the resolvent as in eq. (\ref{resolvent}), the relation to the spectral density changes compared to eq. (\ref{inversion}):
\begin{equation}
R_1(z)=\frac{1}{\pi}\frac{\partial}{\partial z^*} G_N(z)\ ,
\label{inversionC}
\end{equation}
due to $\partial_{z^*}1/z=\delta^{(2)}(z)$ on $\mathbf{C}$. While it is subtle to extract the non-holomorphic part from the expectation value of the ratio of characteristic polynomials two alternative ways exist to formulate the generation of a two-dimensional delta-function. One way is to use so-called replicas which amounts to introduce $k$ extra pairs of complex conjugated quarks in the chiral Lagrangian. Using the relation to the Toda lattice equation, the replica limit can be made well defined as was developed in \cite{KimJacToda}.
The generating function for the density with $k=1$, ${\cal Z}_{\rm echPT}^{[N_f+1+1^*]}$ on the echPT side thus depends on $\hat{\mu}$, which is why the resulting density also depends on it. The second possibility is to use supersymmetry, which will also involve pairs of complex conjugated Bosons \cite{SCS},
\begin{equation}
R_1(z)=-\frac{1}{\pi}\lim_{\kappa\to0}\partial_{z^*}\left(\left.\partial_u\left\langle \frac{\det[(z-D\!\!\!\!/(\mu))(z-D\!\!\!\!/(\mu))^\dag+\kappa^2]}{\det[(u-D\!\!\!\!/(\mu))(u-D\!\!\!\!/(\mu))^\dag+\kappa^2]}\right\rangle_\nu^{\rm QCD}\right|_{u=z}\right)\ ,
\label{susyresmu}
\end{equation}
and both methods have been shown to be equivalent \cite{KimJacequiv}.
 In either case the silver blaze puzzle is resolved. 
We would also like to mention the papers \cite{Bergere,APottier} where averages of ratios of characteristic polynomials and their complex conjugates were evaluated in a purely complex OP framework. The most prominent example for such an average is the distribution of the phase of $D\!\!\!\!/(\mu)$
\begin{equation}
\left\langle \prod_{j=1}^N\frac{z_j+m^2}{z_j^*+m^2}\right\rangle_N^{[N_f]}\ ,
\label{phase}
\end{equation}
and we refer to \cite{KimJacMaria} and \cite{JacTilo} for a detailed discussion including its physical interpretation.

When trying to compare the RMT predictions to QCD lattice simulations the presence of the chemical potential also has a virtue, despite the sign problem: it couples to the second LEC in the chiral Lagrangian, $F_\pi$, in eq. (\ref{ZechPTmu}). Just as for $\mu=0=a$ a fit of the density to lattice data helps to determine $\Sigma$, a two-parameter fit at $\hat{\mu}\neq0$ allows to measure $F_\pi$ in this way. Because for imaginary chemical potential $i\mu$ the Dirac operator spectrum remains real as discussed after eq. (\ref{Dchiralmu}), this was proposed to determine $F_\pi$ using standard lattice simulations \cite{Pouletal}. There is also a corresponding two-matrix model \cite{ADOS} which in the large-$N$ with rescaled $i\mu$ as in eq. (\ref{weaklim}) becomes equal to eq. (\ref{ZechPTmu}), with $\mu^2\to-\mu^2$ (which is of course still convergent). 

Turning back to real chemical potential most comparisons were done within the quenched approximation, starting with \cite{ATilo}, cf. \cite{TiloJacq} for the influence of topology. 
Despite the difficulty of defining individual eigenvalue distributions these were compared with in \cite{ABSW}.
There are many issues that we have not discussed about the application to QCD with chemical potential, apart from methods different from the OP approach mentioned above. 
This includes the different origin of chiral symmetry breaking at $\mu\neq 0$ that is strongly related to the oscillatory behaviour of the unquenched density and we refer to \cite{KimJamesJac}  and to \cite{Kimreview} for a review. Thanks to its analytical solution the  RMT eq. (\ref{ZRMTmu}) has been used as a testing ground for numerical ways to solve the sign problem. This has been successful for the RMT itself and is reported in \cite{Jacsubset}.
 
Furthermore, the study of the singular values of the QCD Dirac operator with chemical potential has been proposed in \cite{TiloTakuya}, that could also be used to analyse different phases (not depicted in Figure \ref{fig:phasediag}) for larger values of $\mu$. Here both the lattice realisation and the RMT have many open questions, and a first step in RMT has been taken in \cite{AStr}. Here a different kernel is found at the origin, the so-called Meijer G-kernel \cite{Arno} that generalises the Bessel kernel. 
The mathematical question of universality in the limit of weak non-Hermiticity for non-Gaussian RMTs is still open at the moment, despite first heuristic attempts \cite{A02u}. For a further discussion including all  known limiting kernels in this weak non-Hermiticity limit resulting from Gaussian models we refer to \cite{APhillips} and references therein.

\section{Summary}

Let us briefly summarise the lessons we have learned in these lectures. The theory of orthogonal polynomials on the real line and in the complex plane and their properties has been extensively discussed. A particular focus was put on weights including characteristic polynomials, as this allowed us to determine the Dirac operator spectrum of QCD including the contribution of $N_f$ quarks flavours in a random matrix approach. Based on the global symmetries of QCD we have taken a limit to  the low energy regime that eventually leads to a random matrix description in a controlled approximation, to which corrections can be computed. Even though this is an unphysical limit it can be compared to numerical solutions of QCD on the lattice. This leads to a determination of the low energy constants of the theory and tests of lattice algorithms to be used then in the physical regime. 
We have shown how the introduction of more details in the theory of QCD such as the effect of finite lattice spacing or of chemical potential, that make the eigenvalues complex, modify the spectrum. And we have mentioned several open problems relevant for RMT and for its application like the computation of new correlation functions or the proof of universality for existing results. They hopefully contribute to keep this an interesting and lively area of research for both mathematicians and theoretical physicists.\\

{\sc Acknowledgements:}
It has been a great pleasure to be back in Les Houches, now as a lecturer, and I would like to express my sincere gratitude to the organisers for inviting me.  
My understanding of this subject would not have been possible without the many collaborations on this topic, and I would like to thank all my coworkers for sharing their insights and compassion. I am particularly thankful to Jac Verbaarschot for many valuable comments on a first draft of this manuscript.
Furthermore I would like to thank the Simons Center for Geometry and Physics at Stony Brook University (programme on Statistical Mechanics and Combinatorics 2016) for the kind hospitality where part of these lecture notes were written up.  This work was partly supported by DFG grant AK 35/2-1.

\end{document}